# Non-Visual Effects of Road Lighting CCT on Driver's Mood, Alertness, Fatigue and Reaction Time: A Comprehensive Neuroergonomic Evaluation Study

Jinchun Wu[a], Zelei Pan[a, b]; Yixuan Liu[a]; Quan Chen[a, b]; Feng Zang[b]; Annette Chabebe[a]; Chengqi Xue[a]*

[a] School of Mechanical Engineering, Southeast University, Suyuan Avenue 79, Nanjing, Jiangsu Province, 211189, P.R.China

[b] Nanjing City Lighting Construction and Operation Group Co. LTD, Huju Road 177, Gulou District, Nanjing, Jiangsu Province, 211189, P.R.China

**Abstract**

Good nighttime road lighting is critical for driving safety. To improve the quality of nighttime road lighting, this study used the triangulation method by fusing "EEG evaluation + subjective evaluation + behavioral evaluation" to qualitatively and quantitatively investigate the response characteristics of different correlated color temperature (CCT) (3500K, 4500K, 5500K, 6500K) on drivers' non-visual indicators (mood, alertness, fatigue and reaction time) under specific driving conditions (monotonous driving; waiting for red light and traffic jam; car-following task). The results showed that the CCT and Task interaction effect is mainly related to individual alertness and reaction time. Individual subjective emotional experience, subjective visual comfort and psychological security are more responsive to changes in CCT than individual mental fatigue and visual fatigue. The subjective and objective evaluation results demonstrated that the EEG evaluation indices used in this study could objectively reflect the response characteristics of various non-visual indicators. The findings also revealed that moderate CCT (4500K) appears to be the most beneficial to drivers in maintaining an ideal state of mind and body during nighttime driving, which is manifested as: good mood experience; it helps drivers maintain a relatively stable level of alterness and to respond quickly to external stimuli; both mental and visual fatigue were relatively low. This study extends nightime road lighting design research from the perspective of non-visual effects by using comprehensive neuroergonomic evaluation methods, and it provides a theoretical and empirical basis for the future development of a humanized urban road lighting design evaluation system.

# 1. Introduction

According to the Global Status Report on Road Safety-2018 (Organization, 2018), global road mortality rates approach 1.35 million deaths per year. According to Chinese road accident statistics, 70% of accidents result in fatalities, posing a serious threat to human life. In addition, according to statistics from the International Commission on Illumination (CIE), it has three times the accident rate for nighttime driving than for daytime driving. Prior studies have also confirmed that more than half of traffic fatalities occur at nighttime compared to daytime, and that good nighttime road lighting can significantly reduce traffic accident rates (Elvik, 1995; Pan et al., 2023). Therefore, the quality of nighttime road lighting environment is a critical factor affecting traffic safety (Schreuder, 1998).

The cognitive process of driving mainly includes three stages: perception, decision, and operation (Baumann & Krems, 2007; Nemeth, 1995). Among these, the quality of perceived visual information influences the driver's ability to make quick and accurate decisions and maneuvers to some extent. Pulat (1997) asserted that 80% of the information a driver takes in while driving comes from the visual channel. Due to the paucity of visual science research in the past, people tended to think that the only purpose of nighttime road lighting was to serve as a light source so that drivers could swiftly detect the visual information around the road and perform the appropriate actions. Consequently, prior studies on road lighting and driving safety mainly focused on the *"impact of lighting conditions on drivers' visual performance"* (Pan et al., 2023), such as different types of street lamps (Fotios & Cheal, 2007), luminance (Lewis, 1999; Plainis et al., 2006), visibility (Mayeur et al., 2010), color temperature (Dong et al., 2017; Q. Zhang et al., 2008) or lighting layout design (Saraiji, 2009) on drivers' visual performance and reaction time. The aforementioned studies have enriched and optimized the content of evaluation standards for road lighting design to some extent, increasing road driving safety.

Notably, in 2001, it was discovered that a type of retinal ganglion cells known as intrinsically photosensitive (ipRGCs) (Brainard et al., 2001; Thapan et al., 2001) was particularly sensitive to blue light at 460 to 490 nm (Bailes & Lucas, 2013; Lockley et al., 2006; Phipps-Nelson et al., 2009). Following the capture of light signals, ipRGCs will carry out neural projection responses in different regions of the brain, creating a neural pathway distinct from vision formation, affecting human hormone secretion, circadian rhythm, mood, alertness and other cognitive processing (H. Li et al., 2017; Rosenthal & Wehr, 1992; Van Bommel, 2006; Webb, 2006). The above-mentioned non-visual effects of light on human physiology and psychology are commonly defined by scholars as non-visual effects.

The discovery of ipRGCs has accelerated non-visual effects research. People are becoming increasingly concerned about the effect of lighting environment on human cognition and other non-visual aspects (Bellia et al., 2011). Over the last two decades, researchers have focused on the effects of indoor lighting conditions on human mood, alertness, work performance and biological rhythm, such as offices (KHADEMAGHA et al., 2015; Zeng et al., 2021), classrooms (Hathaway, 1993; Xiao et al., 2021) and special work space (i.e., hospital operating room, subway dispatching workstation, etc) (Englezou & Michael, 2022; T. Liu, 2020; S. Yang et al., 2023). These studies have revealed the non-visual regulation effect of lighting conditions on human body to some extent, and provided a research method for reference. However, no consensus has been reached on the characteristics and rules of lighting's non-visual effects on individuals. Some representative research conclusions include: blue light rich in short-wavelength characteristics (Cajochen et al., 2005; Chellappa et al., 2011; Mills et al., 2007) or high illumination (Cajochen et al., 2000; Figueiro et al., 2016; Smolders et al., 2012) lighting environment has an inhibitory effect on melatonin, which is conducive to reducing sleepiness and improving human alertness (Viola et al., 2008). Human mood tends to be more positive in the presence of high color temperature or rich blue light (Figueiro et al., 2016; Iskra-Golec et al., 2012). This implies that we could be expected to optimize the quality of the external lighting environment through design, so as to improve the individual's physical and psychological state, as well as work efficiency, satisfaction and happiness.

Hence, in today's rapid economic development, which emphasizes the pursuit of a higher level of better life, the research and design work for nighttime road lighting should also step up to a higher level. However, there are currently few studies on the non-visual effects of road lighting environments on drivers. As the main body of the system, the driver's physiological and psychological states and changes will have a large impact on his or her decision-making and behavior in the overall human-vehicle-environment system (C. Zhang & Eskandarian, 2020), and affect road driving safety. At present, a large number of studies have shown that drivers' mood state, alterness level and fatigue level are important inducements leading to dangerous driving and traffic accidents (Cunningham & Regan, 2016; Lal & Craig, 2001; Scott-Parker, 2017; Steinhauser et al., 2018). At the same time, according to the findings of existing studies on non-visual effects, it can be inferred that the nighttime road lighting environment may have a regulating effect on drivers' non-visual indicators such as mood, alterness and cognitive efficacy. Therefore, on the basis of guaranteeing visual safety, it is necessary to further verify and discuss the non-visual effects of nighttime road lighting environment on drivers, and summarize the response relationship and characteristic rule between individual physiological and psychological indicators and lighting environment, so as to provide a basis for designing and optimizing night road lighting environment and improving night driving safety, health, efficiency and comfort.

Accordingly, the purpose of this study is to investigate the influence of different CCT levels on non-visual indicators (mood, alertness, fatigue, and response time) of individuals performing different driving activities by replicating the nighttime road lighting environment of urban motor vehicles. We focused on analyzing the response relationship between CCT and various non-visual indicators by using the triangulation method of "EEG evaluation + subjective evaluation + behavioral evaluation", as well as screening the appropriate CCT range for nighttime driving. Because the surrounding situation and environment of the driver are always in a state of dynamic change during the real driving process, the driving situation faced and the driving tasks to be performed are not static. Moreover, changes in external driving situations will also alter the driver's emotional state and vigilance level. Therefore, the type of driving task is also taken as a research-independent variable in this study to make the simulated experimental conditions closer to the real situation, so as to explore and verify the changing characteristics of drivers' various physiological and psychological indicators affected by different lighting conditions under different driving task situations. In practice, the results can serve as a theoretical and empirical basis for the future construction of a humanized urban road lighting design evaluation system that incoeporates visual and non-visual effects, so as to improve the night driving experience and provide drivers with a safer, healthier, more comfortable, greener and more efficient humanized lighting environment.

## 2. Literature review

### 2.1 Study on road lighting design based on vision science theory

Research points out that 80% of drivers' access to external information comes from visual channels (Krems & Baumann, 2009; Pulat, 1997). Once the visual function is abnormal, it will directly affect the quality of driving information acquisition, leading drivers to make wrong decisions and behaviors, eventually causing traffic accidents. At present, many studies have explored the visual characteristics of drivers under different road lighting conditions, including light and dark vision (light adaptation and dark adaptation) (Dong et al., 2018; Yi et al., 2012), mesopic vision (Bullough & Rea, 2004; Stockman &

Sharpe, 2006; Vicente et al., 2023; Yong et al., 2011), dynamic or static state visual characteristics (Almeida et al., 2014; He et al., 2017), peripheral vision (He et al., 2020), etc. The preceding research relied heavily on indoor simulation platforms, real road tests, and visual science research experiments. The influences of different lighting indicators (e.g, background brightness, luminance, CCT, visibility, etc.) and lighting parameters on drivers' visual efficacy (e.g, reaction time and accuracy rate), visual recognition distance (Hoseinabadi et al., 2015; Sun et al., 2021), color recognition (Davidovic et al., 2019; Whillans & Allen, 1992) and other aspects were discussed and verified. The above-mentioned research findings provided an important foundation for the development of the current road lighting design standards and evaluation models. The research on road lighting design based on vision science theory has mainly focused on solving the basic visual safety problems of "seeing clearly", "discerning clearly" and "seeing far" while driving at night, so as to make the light environment of nighttime road lighting conform to the visual characteristics of human eyes. However, these studies did not fully consider the impact of lighting environment on other physiological and psychological aspects of drivers.

**2.2 Study related to visual effects of road lighting**

There has been a significant amount of rather mature research on the visual effects of road lighting. The majority of studies have primarily focused on how different lighting factors (e.g., luminance and CCT) affect drivers' visual performance in order to determine the range of lighting parameters with the best visual effects, which can be used to guide the design optimization of road lighting environments and improve nighttime driving safety and visual comfort. For instance, He et al. (2017) investigated the effect of different areas and different brightness environments inside the tunnel on the driver's color discrimination performance of the target object, and proposed the optimization proposal of tunnel lighting design that should consider the signal-to-noise ratio and medium distance brightness of the transition zone and the inner zone respectively. Domenichini et al. (2017) explored the effects of different types of lighting systems (LED vs. HPSV) on tunnel driving performance using a simulated driver. They discovered that the behavioral performance of drivers showed statistically significant differences with the changes of lighting types. The simulated LED lighting environment was more favourable to inducing drivers to perceive dangers in advance and make more timely and effective decisions, resulting in a shorter driving response time and improved driving safety. Liang et al. (2020) investigated the effects of different values of luminance and CCT on human visual performance and reaction time in the interior zone tunnels. They found that increasing CCT could improve visual performance and significantly reduce driving reaction time, particularly at CCT of 5000 K. X. Zhang et al. (2017) proposed three indicators, visual performance, light efficiency, and production cost, to study the visual performance of drivers, when drivers recognized an obstacle in tunnel LED lighting environment with different characteristics. They discovered that increasing the light color rendering can improve the visibility without increasing the light power and provided recommendations of color temperature with color rendering. Dong et al. (2020) examined the effect of different CCTs on perception luminance of human eye under different levels of fog concentration and luminance, and found that the main factors influencing visual perception are luminance and fog concentration and LEDs with higher CCT (6500 K) could provide higher visual clarity. Similarly, in some studies, test subjects were asked to identify the target object in the environment illuminated by LEDs with varying CCTs, and it was discovered that intermediate-to-high CCT could contribute to faster reaction time and lower visual load (Kang et al., 2021; Liu et al., 2013). The aforementioned studies mainly explored the relationship between road lighting indicators and visual effectiveness, and these research findings provided a theoretical and empirical basis for the optimal design of road lighting, meeting the visual needs of night driving while also improving the visual comfort and safety.

**2.3 Study related to non-visual effects of road lighting**

At present, there are relatively few studies both at home and abroad that investigate the impact of road lighting environment on drivers and driving safety from the perspective of non-visual effects, and the relevant research objects are still mainly limited to the tunnel interior lighting environment. For example, Meng et al. (2020) explored the influence of monolor light inside the tunnel on the fatigue state of drivers through simulated driving experiments. They found that the use of red special lighting environment has obvious visual stimulation effect on drivers and can play a visual warning role, which can alleviate the phenomenon of "tunnel hypnosis" and improve driving safety. Peng et al. (2022) investigated the relationship between drivers' physiological and psychological states and lighting environments when they passed through the interior zone of a long tunnel under various lighting conditions through the analysis of breathing rate and heart rate (HRV) in the real car experiment. They discovered that drivers' perceptions of environmental luminance, rather than road luminance, can more accurately reflect their physiological and psychological states in the long tunnel, and that mental stress of drivers decreased more obviously when the background luminance of long tunnel increased. Liu et al. (2021) used simulated driving experiments to compare the driver's reaction time and pupil area difference to the tunnel lighting environment with different short-wave relative spectral values (RSV) and CCT. They propsoed the recommended values of CCT and RSV for LED light sources with different luminance levels in different areas based on the driver's visual performance and non-visual response characteristics in the lighting environment inside the long tunnel. Li et al. (2021) used VR and EEG to compare the effects of different CCTs on the driver's reaction time and visual performance in different driving scenarios. They found that driver's reaction time is fastest at 5000 K, and that low color temperature (2000 K, 3000 K) causes more fatigue than middle-high color temperature (4000 K ~8000 K). It should be noted, however, that if the principle of luminance technology is different, the spectral composition of the light source is different, and thus the non-visual effects of the light source on the human body are also different. Thus, while VR technology can simulate the virtual light environment equivalent to the actual light source at the visual perception level, it is difficult to verify whether the two physical light source parameters are consistent under current technical conditions. As a result, this paper believed that VR technology may be more suitable for studying visual effects and the evaluation of road lighting, while its research on non-visual effects still has limitations in the biological sense.

## 2.4 The use of EEG technology in cognitive research

EEG has attracted increasing interest from researchers assessing factors of human behaviors, especially driving behaviors (L. Yang et al., 2018). As a tool for neuroergonomics, EEG has been regarded as an effective approach for understanding, evaluating, and improving human performance (Jackson & Bolger, 2014; Wu et al., 2023), and it can be used in both laboratory and applied settings. The EEG signal is a representation of the brain's electrical activity recorded from electrodes placed on the scalp. Ergonomists have reported that the cerebral electrical activities are significantly associated with driving performance (Lin et al., 2005). The use of EEG is considered to be sufficient to explore underlying cognitive processes, which are representative of mood (Pan et al., 2023), alterness (Razak et al., 2022), fatigue (X. Li et al., 2021), and workload (Berka et al., 2007).

The EEG spectral components, for instance, alpha ($\alpha$) (8-13Hz), beta ($\beta$) (14-30Hz), delta ($\delta$) (1-4Hz) and theta ($\theta$) (5-7Hz), are used to determine activity levels for different cognitive activities. Specifically, it has been shown that good mood regulation ability is associated with the reduction of spontaneous alpha activation in the left frontal lobe (Brown et al., 2011; Palmiero & Piccardi, 2017; Yoshiike et al., 2018). Beta is closely related with alterness task results (e.g., simulated driving and regular psychological tests) (Herrmann et al., 1979; S.-F. Liang et al., 2006; Makeig & Inlow, 1993), and it is sensitive to changes in driver alertness during driving (Morales et al., 2017). When the subject's alertness is low, the EEG often shows the characteristics of increased slow-wave rhythm and weakened fast-wave rhythm (Lee et al., 2014). Slow waves delta (1-3Hz) and theta (4-7Hz) are associated with fatigue and sleepiness, and a large number of studies have shown that delta and theta rhythms are enhanced while fast wave rhythms (such as beta waves) are weakened under fatigue state (Khushaba et al., 2010). Therefore, a body of previous studies used the ratio of slow wave to fast wave (e.g., *(θ+α)/β, α/β, (θ+α)/(α+β)* and *θ/β*), to characterize the fatigue degree of humans (Bose et al., 2019; Jap et al., 2009; W. Li et al., 2012).

## 3. Methods

### 3.1 Participants

A total of 21 healthy graduate and undergraduate student volunteers (mean age = 23.8, SD = 0.96, 16 males, 5 females) with driver's license were recruited from Southeast University via the campus forum. The average driving age of the subjects was 4.5 years (SD = 1.01). A power analysis using G*power (Version 3.1.9; Faul et al., 2007) indicated that a sample size of at least 12 participants per condition would afford 80% power ($\alpha=0.05$, $f=0.25$). All the participants were right-handed, had normal vision or corrected to normal vision and had no history of neurological or psychiatric illness. One week before the experiment, all subjects were instructed to keep a normal schedule, avoid prolonged exposure to special light environment, get enough sleep (about 8 h) the day before the experiment, and refrain from using any stimulants (i.e., alcohol and caffeine) that might affect their EEG data 48 hours prior the experiment. All participants provided their written informed consent before the experiment and received a monetary compensation of 80 RMB for their participation after the experiment.

### 3.2 Experiment apparatus

The control of variables in the simulated experimental setting and its deviation from the actual night road lighting environment is the key to the reliability and validity of the entire experiment. Therefore, in order to restore the lighting environment of outdoor motor vehicle roads at night in the laboratory as much as possible, this study selected a room with a relatively regular internal structure (6m*9m*3m) in Human Factor Lighting Lab of Nanjing Lighting Group and built a darkroom. The windows around the darkroom are covered with full blackout stickers, supplemented by black thick blackout curtains (curtain material absorbs light to avoid light reflection inside the darkroom), and the surrounding walls and top surfaces are covered with black light-absorbing flocking cloth to avoid external light sources pollute. In addition, the laboratory ground is surfaced with asphalt slabs to imitate as closely as possible the light reflection circumstances of the real pavement material. The experiment was carried out using a driving system comprised of a driving simulator and driving scene design software. The hardware part of the driving simulator consists of three display screens (Samsung, C34G55TWWC, 4K-165Hz), which can provide drivers with a viewing angle of about 180° horizontally; an adjustable Half-bucket seat (XDracing); a steering wheel and a pedal (Logitech, G923 TRUEFORCE), the average sensitivity of the device is set to 80%. On the right side of the simulated driving seat is a laptop that is used to run the Psychomotor Vigilance Task (PVT) programme built by Eprime2.0 and record participant reaction data. To measure the participants' critical flicker frequency (CFF), a flicker fusion frequency metre (Beijing tianqiao instruments, DB-II-118) was placed on the left side of the simulated driving seat. Behind the driver's seat, the Brain Vision actiCHamp EEG device was installed. Brain activity was recorded using 64 Ag/AgCl impedance-optimized electrodes (Brain Products, Munich, Germany), referenced to the FCZ electrode, sampled at 1000 Hz and wide band filtered (0.5–100 Hz), placed according to the international 10-20 system. The layout of the drakroom is as shown in Fig. 1.

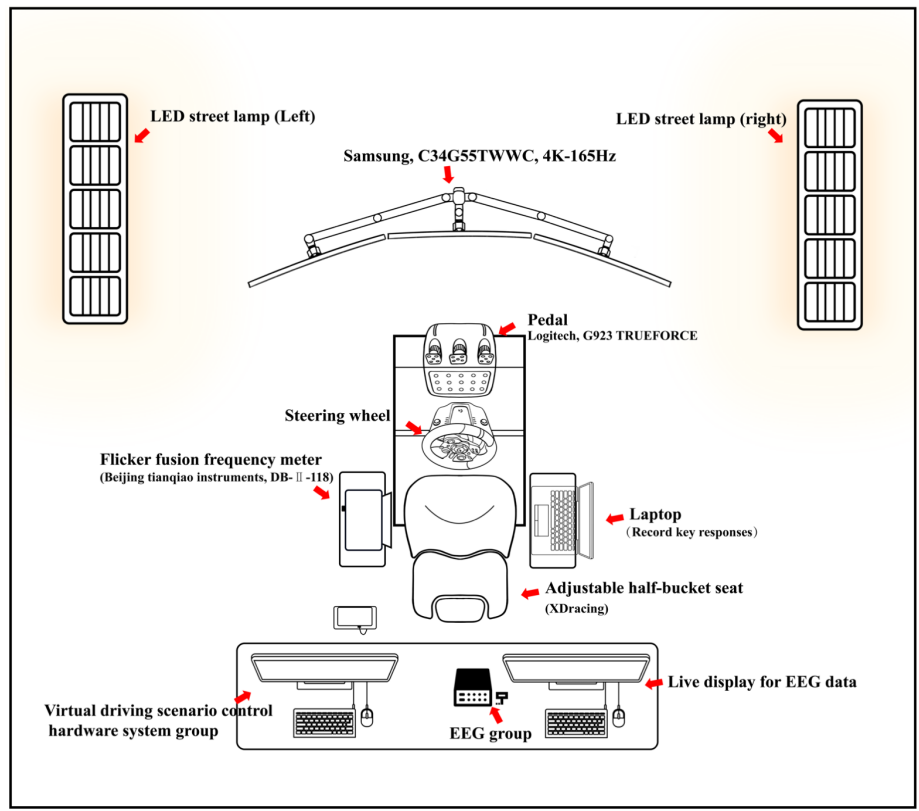

**Fig. 1**. A top view of the interior layout of the darkroom for the light environment simulation experiment.

## 3.3 Light manipulation

When studying the non-visual effects of road lighting environments, the visual effects of lighting must be considered because cone cells and rod cells may regulate the non-visual effects of ipRGCs on the human body. Thus, in this study, the display screen was used to present the visual image information of the virtual night driving scene. In addition, it is important to note that the road lights in the virtual driving scene should match the CCT and illuminance of the lamp light source in the darkroom in the visual perception layer. The road lighting environment in the virtual driving scene should be controlled as much as possible to give people the same visual feeling in terms of color and brightness, so as to avoid the adverse impact of visual perception differences on the accuracy of the experimental results. Therefore, pre-experiments with spectrometers measured the light color values of external luminaries' CCTs (2500k, 3500K, 4000k, 4500K, 5000k, 5500K and 6500K). A corresponding floodlight model with the same RGB values (see Table 1) is created in 3D Max software to ensure that the light environment in the simulated driving scene is equivalent to that provided by the actual lamps in the visual color perception layer. Finally, the lighting model in 3D Max was imported into the SCANeR studio software, and then the brightness value of the ambient light inside the virtual scene was manually fine-tuned, so that the light environment in the virtual scene was consistent with the subjective visual brightness provided by the light environment provided by external lamps. In this study, the screen display brightness was set to 10% to control and minimise the display light source's effect on the experimental results. Photometric measurements showed that the spectral power distribution (about 440nm~630nm) of the mixed light jointly provided by the screen and the lamp is less different than that of the single light source provided by the lamp. Studies have shown that ipRGCs are mainly sensitive to blue light at around 490nm. Since the spectral composition of the hybrid light source and the single luminance light source are similar in this band, it can be inferred that the above two light sources have similar stimulating effects on ipRGCs and the induced non-visual effects are also similar.

**Table 1**. The measured data of the color temperature of the light source of external lamps and the corresponding light color value

| Items | Parameters | | | | | | |
|---|---|---|---|---|---|---|---|
| CCT | 2500K | 3500K | 4000K | 4500K | 5000K | 5500K | 6500K |
| RGB | (255, 161, 72) | (255, 196, 137) | (255, 209, 163) | (255, 219, 186) | (255, 228, 206) | (255, 236, 224) | (255, 249, 253) |

In addition to the screens displaying the virtual driving scene in the darkroom, all other luminous screens or equipment (brightness adjusted to 3%) are covered with black flocking cloth to prevent additional light sources from polluting the experimental stimulation light environment. The simulated light source of road lamps within the darkroom is provided by Nanjing Lighting Group's specially customized LED street lamp specification intelligent lighting equipment, which is composed of control modules, lamps, lampshades, and brackets. The dimming control module (see Fig. 2) can remotely control the CCT (3000K-6500K) and illuminance (0-100%) of a single lamp.

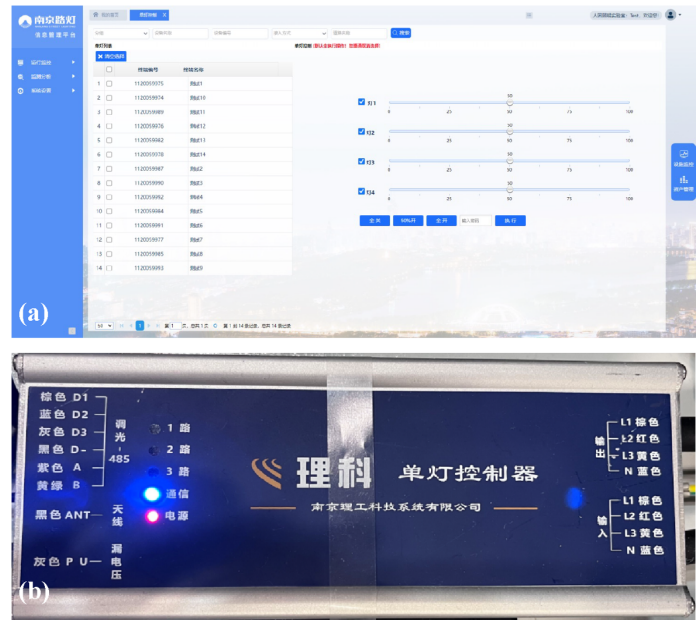

**Fig. 2**. (a) shows a screen shot of dimming system software (single light control); (b) is the LED road lamp intelligent control dimming hardware module.

## 3.4 Experimental Tasks

In this study, three kinds of daily driving task conditions (monotonous driving; waiting for red light and traffic jam; car-following task) were selected as the design prototypes of the driving simulation experiment task types, so as to mimic the driving situation as naturally and realistically as feasible. And the non-visual effects of different road lamp CCTs on drivers' mood, alertness and fatigue can be compared and evaluated. It should be noted that during the whole simulation driving process, participants were only required to control the simulated vehicle on a straight lane, and the surrounding traffic flow density is set as $\rho=0.6$ through the software.

**Task 1 (base task- monotonous driving)**. For task 1, no abnormal stimuli (such as traffic lights, obstacles, etc.) were set to simulate the monotonous driving situation under normal circumstances (see Fig. 3a). Participants were required to maintain a state of natural relaxation, control the virtual vehicle at a speed of 60km/h, obey traffic rules and keep the lane unchanged throughout the process. Task 1 was regarded as the baseline task, and the participants' mood, alertness and fatigue were measured and recorded under the stable and natural state.

**Task 2 (waiting for red light and traffic jam)**. Task 2 simulated the situation of frequently encountering red lights and randomly encountering traffic jams while driving (see Fig. 3b). In previous studies, such task scenarios were mainly used to induce negative emotions in drivers. However, this experiment is mainly aimed at simulating and restoring this daily driving situation, and truly recording and comparing the changing characteristics of the driver's mood and cognitive functions under different color temperature levels. According to the research results, we could evaluate and select the lighting scheme that is most beneficial to the driver to maintain a good physical and mental state and ensure driving safety. At this stage, the vehicle speed limit is also 60km/h.

**Task 3 (car-following task)**. The dual-task experimental paradigm was used to design the the task 3 (primary task: car-following task; secondary task: auditory PVT task). In this stage, subjects are required to control the simulated vehicle to follow the target vehicle (a red truck, see Fig. 3c) at a subjective safe distance. If the target car changes lanes, the participant shall change lanes as well, and no overtaking or rear-end collision are permitted. This task did not have any specific speed constraints. While juggling the car-following task, participants were required to respond as quickly as possible to a random sound stimulus (a simulated ring tone that lasted for 3s) appeared randomly (50 to 70s between stimuli) by pressing a key (the "space bar"). The PVT task reaction time was used to represent the driver's sustained attention and cognitive control level, so as to evaluate driving safety in different light settings.

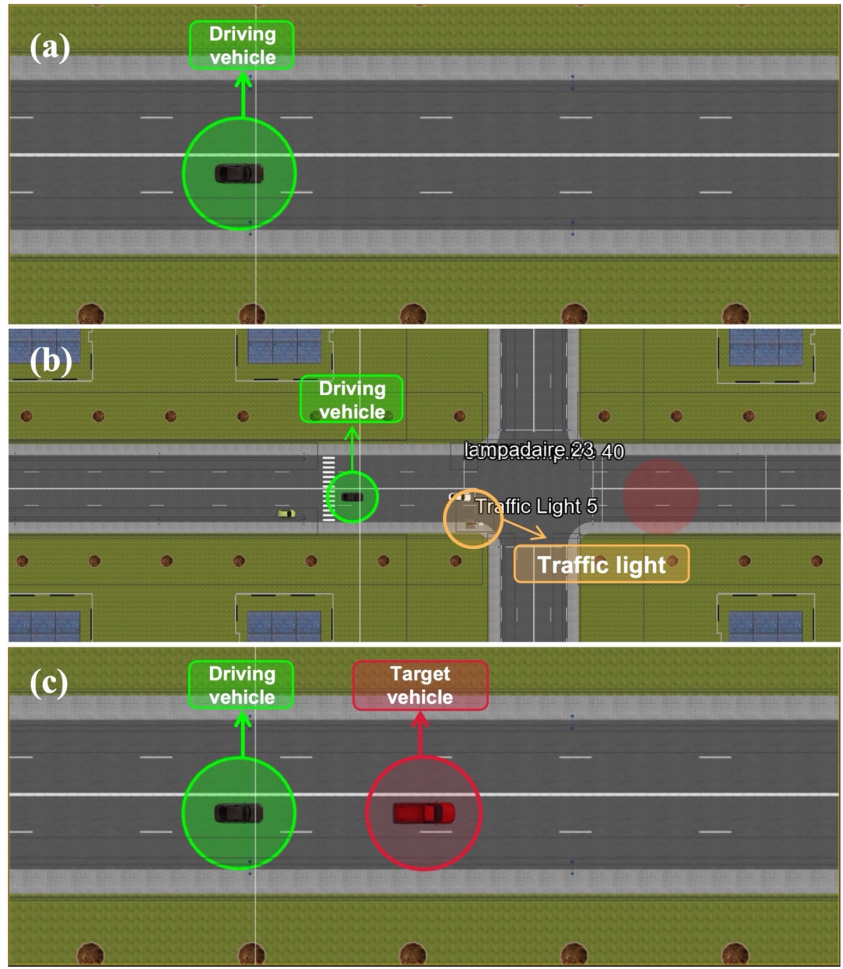

Fig. 3. Picture of the software settings of simulation scenarios for thereee tasks. (a) depicts the task 1, monotonous driving, (b) depicts the task 2, waiting for red light and traffic jam; and (c) depicts the task 3, car-following task.

## 3.5 Experimental design and procedures

The experiment's CCT settings were primarily based on the CCT range of road lamps used by national urban motor vehicles (2500K–6500K) (main and secondary arterial roads). According to the actual road measurements, the primary distribution range of the CCTs for motor vehicle road lamps in Nanjing is approximately 3000K ~6000K. Therefore, four CCT levels (3500K, 4500K, 5500K and 6500K) were chosen for this experiment in accordance with the actual setting range of road lighting CCT and the results of prior research. We used the within-subject repeated-measures experimental design with 4 (CCT: 3500K vs. 4500K vs. 5500K vs. 6500K) × 3 (Driving task: task 1 vs. task 2 vs. task 3). Each participant took part in four experiments in succession separated by one day (experiment washout period: 1 day). Latin square experimental design was used to balance results and prevent interference from the sequence of experimental sessions. Each participant took part in the lab experiment for a fixed time (19:00-20:30).

The experiment was conducted from May to August 2022. It should be noted that the illumination factor served as the control variable in this experiment (average ground illumination: 30lx, measured eye-level illumination: about 3lx). The setting of illumination parameters is referred to CJJ45-2015 Urban Road Lighting Design Standard (see 3.3 Standard Value of Motor Vehicle Road Lighting for details).

This study was conducted in the Human Factor Lighting Lab of Nanjing Lighting Group, where external interferences (e.g., lighting, temperature (25°C), humidity (45%) and noise) were strictly controlled. The participants were asked to sit in the driving simulator comfortably at 1500 mm away from the displays. The driving scenes were displayed on three 27-inch LCD monitors (Samsung, C34G55TWWC, 4K-165Hz) with a a resolution of 1920*1080 pixels. The experimental driving tasks were programmed through SACNeR studio software, which could provide participants with immersive virtual images of nighttime driving scenes.

Participants first filled out a form with their personal information (e.g., age and driving experience) before reading an experiment introduction. After the preparation of electrodes (approximately 15 min), participants had a 20-min practice session with all three task conditions. After practice period, the participants put on the full-blackout eye mask, closed their eyes, relaxed naturally, and adapted to the dark environment to eliminate the light history's effect on the experimental results. It should be emphasized that the luminance stability of lamps will be affected if the lamp is opened for too short a time. Therefore, this experiment used full-blackout eye masks instead of turning off the lamps. Followed dark adaptation for 10 min, participants underwent bright adaptation to the current experimental light environment (randomized CCTs). After bright adaptation (3 min), they were asked to perform three randomly organized driving tasks (task 1, task 2 and task 3). After each task, the participants were asked to complete the CFF measurements (take 3 consecutive measurements and average the results) (Landis, 1954; Skrandies, 1985). Subsequently, they were required to finish the following rating scales：

（1）**Subjective emotion rating scale:** This scale combined the Likert 9-point scale with the Self-Assessment Manikin (SAM) (Lang, 2005), where '0' represents the lowest level and '8' represents the highest level.

（2）**Subjective alterness rating scale:** This sacle was adapted from the KarolinskaSleepiness Scale (KSS) (Shahid et al., 2012). Since the driver did not appear extremely sleepy during the actual experiment and could not stay awake at all, the 10th evaluation item in the original KSS scale was deleted in this study.

（3）**Subjective visual fatigue evaluation scale:** Indexes (dry eyed, eyeache, photophobia, hazy, pain in other parts of the body) of this 5-point scale were selected from the classical subjective fatigue evaluation questionnaire (Kuze & Ukai, 2008; Wang et al., 2015), where '0' represents the none level and '4' represents the severe level.

After completing rating scales of each task, particiopants have a 5-min break. After finishing all the task, participants were asked to rate the current experimental light environment using the modified lighting evaluation scale based on the rating scale of Flynn et al. (1973). The Likert 7-point scale has 6 dimensions and consists of 6 pairs of opposite meanings (harsh - soft, dim - bright, insecure - secure, fuzzy - clear, very confusing - not confusing at all, insecure – secure, not satisfied - satisfied). The whole single experiment lasted approximately 1.5 h.

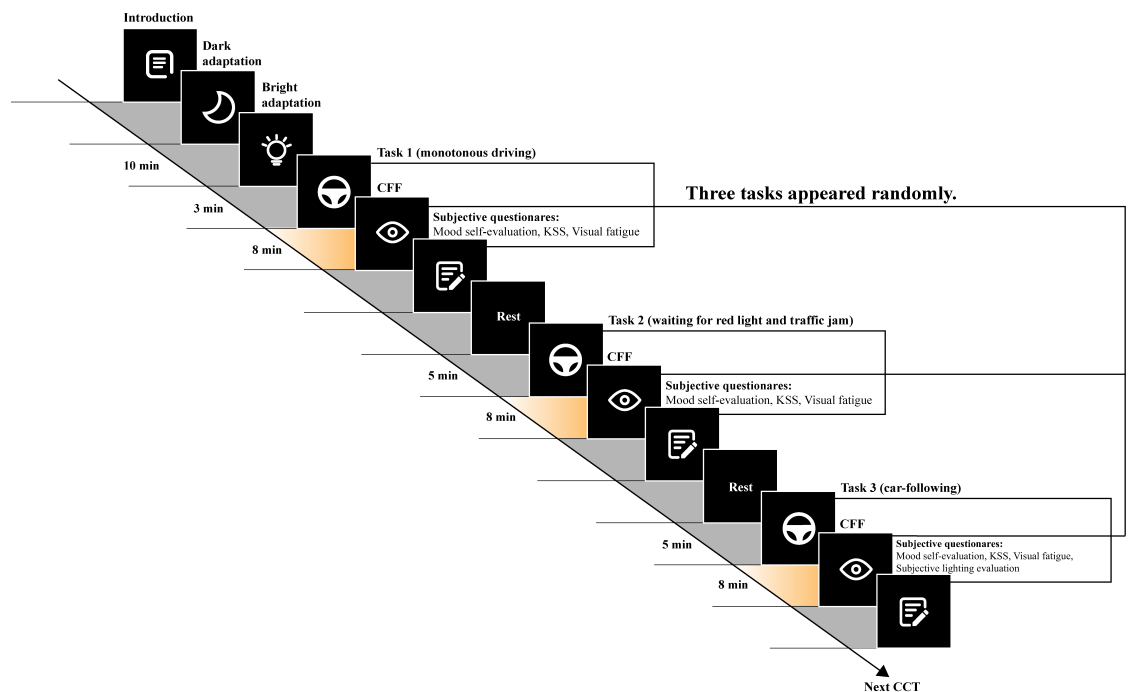

**Fig. 4.** Schematic time course of the experimental procedure.

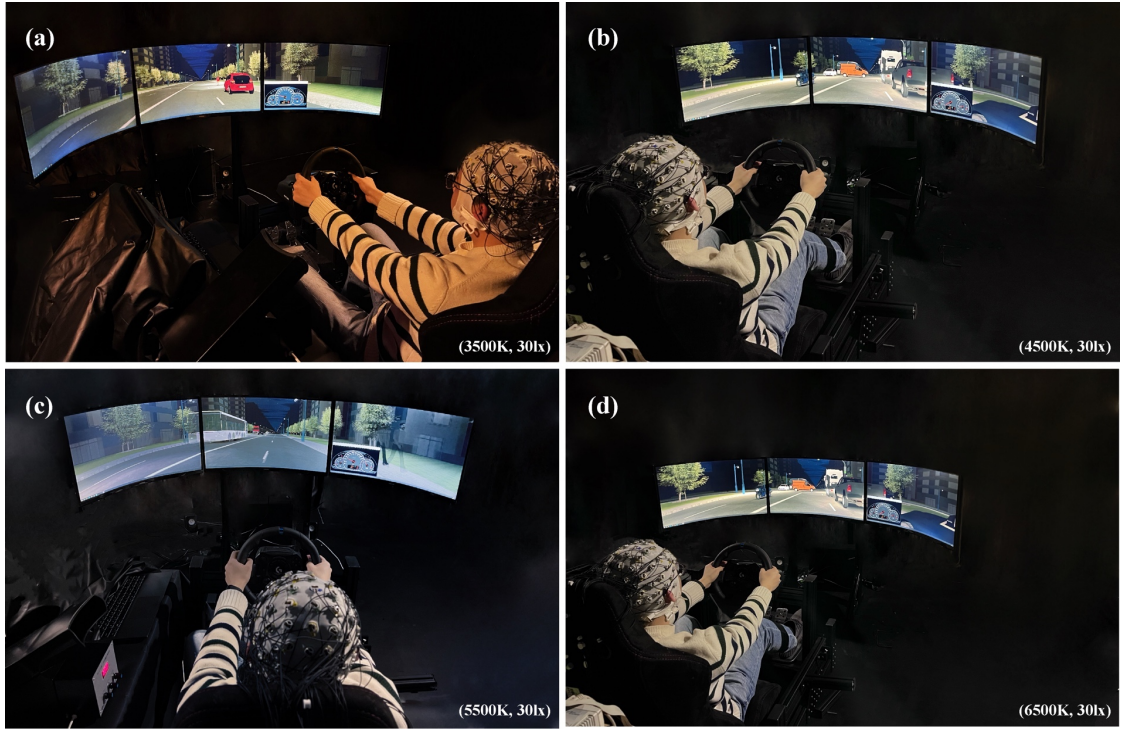

**Fig. 5.** Pictures of the experimental scene. (a) shows the experimental scene with CCT (3500K) and illuminance (30lx); (b) shows the experimental scene with CCT (4500K) and illuminance (30lx); (c) shows the experimental scene with CCT (5500K) and illuminance (30lx); (d) shows the experimental scene with CCT (6500K) and illuminance (30lx).

### 3.6 Data analysis

In total, nine parameters, that is, subjective mood score (*SM*), KSS perfromance, subjective visual fatigue, lighting evaluation score, CFF performance, reaction time, fro-alpha power ($\alpha_{(L/R)}$), parietal-occiptal beta power ($\beta$), frontal-occipital theta and beta power ($\theta/\beta$).

*SM* is defined as the (positive/negative) bias of subjective mood, which is used to represent the dominance of participants' subjective negative emotions against positive emotions while driving. It can be measured according to:

$$SM = \bar{N}_{ij} - \bar{P}_{ij} \quad (i = 1, 2, 3; j = 1, 2, \cdots, 21) \quad (1)$$

Where $\bar{N}_{ij}$ is the mean subjective score of the subjects' negative emotion, $\bar{P}_{ij}$ refers to the mean subjective score of the subjects' positive emotion.

Fro-alpha power $\alpha_{(L/R)}$ is defined as the objective index for mood evaluation, which is the ratio of alpha-band power spectrum values of left brain to right brain. This study focused on the analysis of α waves recorded by the electrode channels of the left and right prefrontal cortex (FP1-FP1, AF3-AF4, F3-F4). $\alpha_{(L/R)}$ is calculated as follows:

$$\alpha_{(L/R)} = \alpha_{ij(L)} / \alpha_{ij(R)} \quad (i = 1, 2, 3; j = 1, 2, \cdots, 21) \quad (2)$$

Where $\alpha_{ij(L)}$ and $\alpha_{ij(R)}$ refer to the alpha power of left hemisphere and right hemisphere, respectively. The value of $\alpha_{ij(L/R)} > 1$ indicates negative bias, whereas indicates positive bias.

Parietal-occiptal beta ($\beta$) power was used as the objective indicator for assessing driver's alertness. The present study concentrated on the analysis of parietal-occiptal beta power, extracted by averaging the beta power from six electrodes (P3, P4, PO3, PO4, O1, O2).

As described above (Section 2.4), previous research often used the ratio of slow wave to fast wave to characterize human fatigue degree. Considering that subjects need to try to stay awake in the simulated driving experiment without sleepiness, this study focused on the analysis of theta and beta waves in the frontal (F3-F4) and occipital (Oz, O1-O2) electrodes, and selected the ratio of theta to beta ($\theta/\beta$) as the fatigue evaluation index.

### 3.6.1 EEG analysis

EEG data were preprocessed offline in MATLAB 2013b (MathWorks Inc.; https://www.mathworks.com) using the EEGLAB toolbox (Swartz Center for Computational Neuroscience, UCSD; http://sccn.ucsd.edu/eeglab). EEG recordings were rereferenced to the average of the TP9 and TP10 channels which were positioned on the left and right mastoids, respectively. Then, the sampling rate of EEG signals were down-sampled to 500 Hz and followed by the use of a 40-Hz low pass and a 0.1-Hz high pass by using the basic finite impulse response filter, respectively. The independent component analysis (ICA) approach implemented in EEGLAB toolbox was utilized to remove the electrooculogram (EOG) and movement-related artifacts according to the topography and temporal activities (Delorme & Makeig, 2004). The EEG data were then segmented into 2-s epochs. To ensure the reliability and validity of the results, only trials with at least 2s artifact-free EEG signals during the task periods were used for further analysis. Subsequently, a fast Fourier transform (FFT) was performed to obtain band power values of the EEG signals.

### 3.6.2 Statistical analysis

We performed statistical analysis with SPSS 26.0. A two-way ANOVA was used to test the significance of the differences in these parameters. Shapiro-Wilk and Levene tests were used to test the normality and homogeneity of variance respectively, and the test results showed that the experimental data met the assumptions of analysis of variance. To prevent the violations of sphericity assumptions, the Greenhouse-Geisser and Bonferroni correction methods were employed whenever necessary. An $\alpha$ level of 0.05 was set to determine the stastical significance.

# 4. Results

## 4.1 Effects of CCT on moods

***Subjective ratings on moods:*** ANOVA results of *SM* (see Table 1) showed significant main effects of CCT (F (3, 60) = 7.639, p < 0.001, $\eta_p^2$= 0.276), and Task (F (2, 40) = 18.622, p < 0.001, $\eta_p^2$= 0.482). However, there was no significant interaction between CCT and Task (F (4.136, 82.719) =0.633, p = 0.645, $\eta_p^2$=0.031). Pairwise comparisons revealed that the *SM* of 6500K was higher than that of 3500K (△=1.272, SE = 0.329, p < 0.01) and 4500K (△=1.360, SE = 0.353, p < 0.01).

***EEG results on mood ($\alpha_{(L/R)}$):*** No significant main effects were revealed for CCT (F (2.129, 42.576) = 2.284, p = 0.111, $\eta_p^2$ = 0.102), and Task (F (2, 40) = 1.249, p = 0.298, $\eta_p^2$= 0.059). And there was no significant interaction between CCT and Task (F (6, 120) = 0.774, p = 0.592, $\eta_p^2$=0.037).

## 4.2 Effects of CCT on alterness

***KSS scale evaluation results:*** The results of KSS performance (see Table 1) showed significant main effects of CCT (F (3, 60) = 5.588, p = 0.002, $\eta_p^2$ = 0.218), and Task (F (2, 40) = 3.818, p = 0.03, $\eta_p^2$= 0.160). There was also a significant interaction between CCT and Task (F (6, 120) = 2.288, p = 0.04, $\eta_p^2$=0.103). CCT simple effects were significant for task 1 (F (2.184, 43.683) = 7.554, p = 0.001, $\eta_p^2$= 0.274), and task 2 (F (2.081, 41.628) = 4.207, p = 0.02, $\eta_p^2$= 0.174). For task 1, participants' KSS performance of 3500K was significantly larger than that of 5500K (△=1.667, SE = 0.311, p < 0.001), and the KSS performance of 5500K was significantly decreased than that of 6500K (△=0.667, SE = 0.199, p = 0.019). For task 2, the KSS performance of 4500K was significantly lower than that of 6500K (△=0.952, SE = 0.327, p = 0.052≈0.05). Task simple effects were significant for 3500K (F (2,40) = 4.893, p=0.01, $\eta_p^2$=0.197). Participants' KSS performance of task 1 was significant higher than that of task 3 (△=1.048, SE = 0.312, p = 0.009).

***EEG results on alterness:*** Analysis of the EEG results (see Table 1) on alterness revealed a significant main effect of CCT (F (3, 60) = 3.234, p = 0.028, $\eta_p^2$= 0.139), with mean beta power being lower for 3500K than for 4500K (△=0.079, SE = 0.026, p = 0.037). No main effect of Task arrived at significance (F (2, 40) = 0.774, p = 0.468, $\eta_p^2$= 0.037), and there was no significant interaction between CCT and Task (F (2.629, 52.576) = 0.325, p = 0.781, $\eta_p^2$= 0.016).

## 4.3 Effects of CCT on fatigue

***Subjective visual fatigue evaluation:*** The results of subjective visual fatigue (see Table 1) revealed that there were no significant main effects of CCT (F (3, 60) = 2.066, p = 0.114, $\eta_p^2$= 0.094), and Task (F (2, 40) = 2.945, p = 0.064, $\eta_p^2$= 0.128). And there was no significant interaction between CCT and Task (F (3.259, 65.183) = 0.322, p = 0.825, $\eta_p^2$= 0.016).

*CFF results on fatigue:* The results of CFF (see Table 1) revealed that there were no significant main effects of CCT (F (2.392, 47.848) = 1.176, p = 0.323, $\eta_p^2$= 0.056), and Task (F (2, 40) = 1.854, p = 0.17, $\eta_p^2$= 0.085). And there was no significant interaction between CCT and Task (F (6, 120) = 0.181, p = 0.982, $\eta_p^2$= 0.009).

*EEG results on fatigue:* ANOVA results of the ratio of theta to beta ($\theta/\beta$) (see Table 1) showed no significant main effects of CCT (F (3, 60) = 1.785, p = 0.160, $\eta_p^2$= 0.082), and Task (F (2, 40) = 0.055, p = 0.947, $\eta_p^2$= 0.034). Also, no significant interaction between CCT and Task was observed (F (2.912, 58.248) = 0.714, p = 0.543, $\eta_p^2$= 0.034).

## 4.4 Effects of CCT on drivers' reaction time (RT)

Aalysis of the PVT reaction time (see Table 1) showed a significant main effect of CCT (F (3, 60) = 3.063, p = 0.035, $\eta_p^2$= 0.133). Pair comparisons revealed that the RT of 4500K (M = 1694.482 ms, SE = 316.701) and 6500K (M = 1689.937ms, SE = 296.056) was significantly lower than that of 3500K (M = 1798.998 ms, SE = 312.269) and 5500K (M = 1726.515 ms, SE = 328.148).

## 4.5 Results of subjective lighting evaluation

*Subjective visual comfort (harsh – soft):* As shown in Table 2, the mean subjective visual comfort scores showed a significant main effect of CCT (F (3, 60) = 27.604, p < 0.001, $\eta_p^2$= 0.580). Pairwise comparisons revealed that the mean visual comfort score was higher for 3500K than that for both 5500K ($\triangle$=1.381, SE = 0.223, p < 0.001) and 6500K ($\triangle$=1.905, SE = 0.257, p < 0.001). And the mean visual comfort score was higher for 4500K than for both 5500K ($\triangle$=0.857, SE = 0.186, p = 0.001) and 6500K ($\triangle$=1.381, SE = 0.263, p < 0.001). There was no significance between 3500K and 4500K ($\triangle$=0.524, SE = 0.225, p = 0.183).

*Subjective brightness and perception (dim – bright):* No significant main effect of CCT was observed (F (3, 60) = 0.816, p = 0.490, $\eta_p^2$= 0.580).

*Subjective color discrimination ability (very confusing - not confusing at all):* No main effect of CCT achieved significance (F (1.861, 37.225) = 0.513, p = 0.590, $\eta_p^2$= 0.025).

*Subjective visual clarity (fuzzy – clear):* No main effect of CCT arrived at significance (F (3, 60) = 0.200, p = 0.896, $\eta_p^2$= 0.010).

*Subjective sense of security (insecure – secure):* Statistical analysis revealed that the mean score of subjective feeling of security revealed a significant main effect of CCT ($F_{(2.104, 42.078)} = 7.819$, $p = 0.001$, $\eta_p^2 = 0.281$). Pairwise comparisons revealed that the mean score of subjective sense of security was higher for 3500K than for both 5500K ($\triangle=1.000$, SE = 0.309, $p = 0.025$) and 6500K ($\triangle=1.286$, SE = 0.197, $p < 0.001$).

***Subjective satisfaction (not satisfied – satisfied):*** Analysis of the averged subjective satisfaction score demonstrated that there was no significant main effect of CCT ($F_{(3, 60)} = 7.819$, $p = 0.220$, $\eta_p^2=0.070$).

**Table 1**

Comparison of the means (standard deviations) of different variables from the experiment for all twelve experimental conditions (values calculated from 21 participants).

|  | Variables | 3500K | | | 4500K | | | 5500K | | | 6500K | | |
| --- | --- | --- | --- | --- | --- | --- | --- | --- | --- | --- | --- | --- | --- |
|  |  | Base | Task 2 | Task 3 | Base | Task 2 | Task 3 | Base | Task 2 | Task 3 | Base | Task 2 | Task 3 |
| Mood | $SM$ | -0.238 (2.874) | 3.080 (1.514) | 1.278 (3.828) | -0.533 (2.569) | 2.807 (2.101) | 1.579 (3.905) | -0.262 (3.505) | 3.621 (1.737) | 2.671 (3.016) | 0.348 (3.196) | 4.293 (1.704) | 3.294 (3.710) |
|  | $\alpha_{(L/R)}$ | 1.042 (0.200) | 0.987 (0.066) | 0.973 (0.127) | 0.968 (0.138) | 0.993 (0.130) | 0.956 (0.085) | 0.985 (0.154) | 0.977 (0.053) | 0.971 (0.087) | 1.036 (0.171) | 1.025 (0.149) | 1.019 (0.145) |
| Alterness | KSS | 4.480 (1.692) | 3.710 (1.707) | 3.430 (1.469) | 3.380 (1.658) | 2.670 (0.658) | 2.710 (1.488) | 2.810 (0.873) | 3.050 (0.973) | 3.050 (1.161) | 3.480 (1.327) | 3.620 (1.658) | 3.190 (1.436) |
|  | $\beta$ | 0.251 (0.138) | 0.301 (0.256) | 0.294 (0.267) | 0.325 (0.230) | 0.367 (0.349) | 0.388 (0.287) | 0.338 (0.307) | 0.332 (0.318) | 0.338 (0.416) | 0.318 (0.256) | 0.343 (0.274) | 0.325 (0.245) |
| Fatigue | Subjective visual fatigue | 1.495 (0.398) | 1.448 (0.334) | 1.543 (0.430) | 1.524 (0.360) | 1.486 (0.428) | 1.552 (0.442) | 1.581 (0.400) | 1.576 (0.405) | 1.619 (0.447) | 1.619 (0.490) | 1.643 (0.420) | 1.781 (0.433) |
|  | CFF | 27.890 (3.543) | 28.576 (4.037) | 28.238 (4.142) | 28.438 (4.121) | 28.848 (4.012) | 28.548 (3.630) | 27.571 (3.145) | 28.095 (3.807) | 28.195 (4.172) | 27.352 (4.028) | 28.052 (3.984) | 28.205 (3.390) |
|  | $\theta/\beta$ | 3.994 (2.087) | 3.314 (1.946) | 3.465 (1.532) | 3.354 (2.009) | 3.287 (2.131) | 3.134 (1.856) | 3.851 (2.237) | 4.154 (3.565) | 3.915 (1.963) | 3.272 (1.760) | 3.722 (2.033) | 3.697 (2.470) |
| PVT | Reaction time (ms) | - | - | 1798.998 (312.269) | - | - | 1694.482 (316.701) | - | - | 1726.515 (328.148) | - | - | 1689.937 (296.056) |

**Table 2**

Comparison of the means (standard deviations) of different variables from the six items of lighting evaluation scale for four CCT conditions.

| Dimensions | Adjectives | 3500K | 4500K | 5500K | 6500K |
| --- | --- | --- | --- | --- | --- |
| Subjective visual comfort | harsh - soft | 4.667 (0.658) | 4.143 (0.655) | 3.286 (0.717) | 2.762 (0.889) |
| Subjective brightness perception | dim - bright | 3.330 (1.197) | 3.430 (1.028) | 3.670 (1.197) | 3.810 (1.401) |
| Subjective color discrimination ability | very confusing - not confusing at all | 4.286 (1.146) | 4.619 (0.973) | 4.571 (1.076) | 4.333 (1.354) |
| Subjective visual clarity | fuzzy - clear | 4.238 (0.831) | 4.333 (1.317) | 4.286 (1.056) | 4.095 (1.136) |
| Subjective sense of security | insecure - secure | 4.429 (0.811) | 3.524 (1.078) | 3.429 (1.028) | 3.143 (0.910) |
| Subjective satisfaction | not satisfied - satisfied | 4.000 (1.049) | 4.095 (0.995) | 4.048 (1.203) | 3.476 (0.928) |

# 5. Discussion
## 5.1 The effect of "CCT x Task" on driver's mood

EEG results on mood revealed that the main effect of CCT and the interaction effect of "CCT * Task" were not statistically significant, indicating that there was no significant difference in the influence of external lighting color temperature environment on drivers' mood under different driving task situations. As shown in Fig. 6, $\alpha_{(L/R)}$ in each task stage generally declined first and subsequently increased with the rise of CCT. The values of $\alpha_{(L/R)}$ at 3500K (low), 4500K (middle) and 5500K (high) were all lower than 1 ($\alpha_{(L/R)} < 1$) except for task 1 (base). This indicated that 3500K (low), 4500K (middle) and 5500K (high) CCT lighting environments are beneficial for drivers to maintain a positive mood, but if the monotonous driving task (like task 1) is carried out continuously under low CCT environment, the drivers will have bad mood, and excessively high CCT (6500K) is also unfavorable to the mood experience of drivers. This result was consistent with the findings of Pachito et al. (2018). This may be due to the low composition of short-wave blue light of low-color warm light, which has low activation degree to ipRGCs cells and biological rhythm system (especially melatonin). Thus, it is easy to induce relaxation and low alertness (supported by the experimental results in part 4.2) when performing the originally boring driving tasks, and even cause sleepiness (Rahm & Johansson, 2018; Vetter et al., 2011), resulting in bad moods such as boredom and irritability (supported by the results of the subjective emotion evaluation scale). And, if the CCT is too high (6500K), the driver will experience visual fatigue, discomfort, and psychological insecurity (as demonstrated by the experimental results in parts 4.3 and 4.5).

Although in this experiment, the interaction of "CCT x Task" has not been found to have a significant impact on the driver's mood, but because this experiment did not consider the influence of the "luminance" factor on the driver's mood. Therefore, it can be speculated that, compared with the CCT factor of the external lighting environment, individual ipRGCs received relatively stronger illumination stimulus (Ru et al., 2019) (control ground average illumination: 30lx, which is the highest threshold of illumination specified in Urban Road Lighting Sesign Standards), resulting in the activation degree of their prefrontal emotional brain region in a relatively stable state. Therefore, no matter how the external CCT environment changed, the difference of prefrontal $\alpha_{(L/R)}$ was not significant. As a result, on the basis of this experiment, it is necessary to consider the moderating effect of illumination factor on drivers' mood, as well as conduct further experimental research on the non-visual effects of the interaction of "CCT x Illumination x Task" on drivers. The results of subjective emotion evaluation also showed that the interaction of "CCT x Task" was not statistically significant, but the main effect of CCT was marginal significant. As can be seen from Fig. 7, *SM* showed an overall increasing trend, and is lowest at the middle CCT level (4500K), indicating that participants may believe that driving at the middle CCT has the best mood experience, which is basically consistent with the results of EEG evaluation analysis.

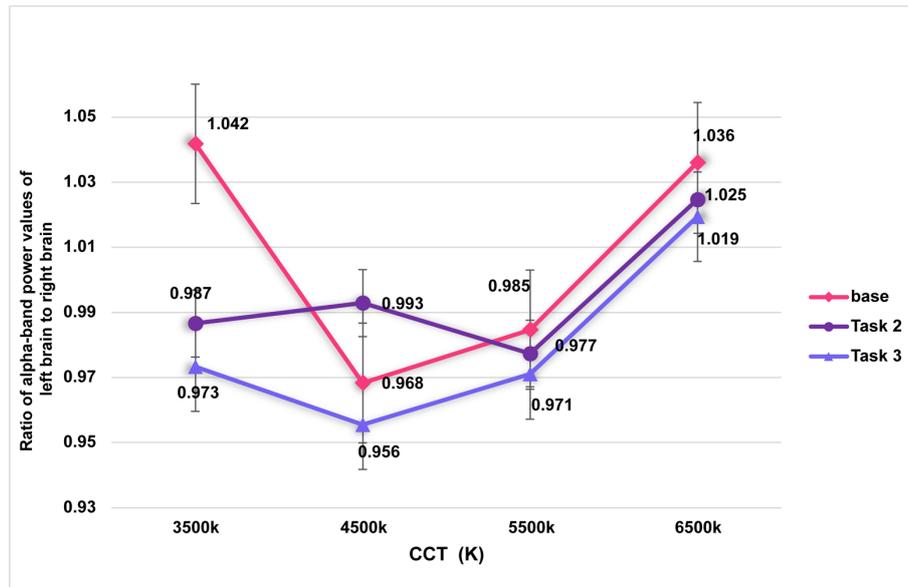

**Fig. 6**. Line chart of change of $\alpha_{(L/R)}$ at various CCT and task level.

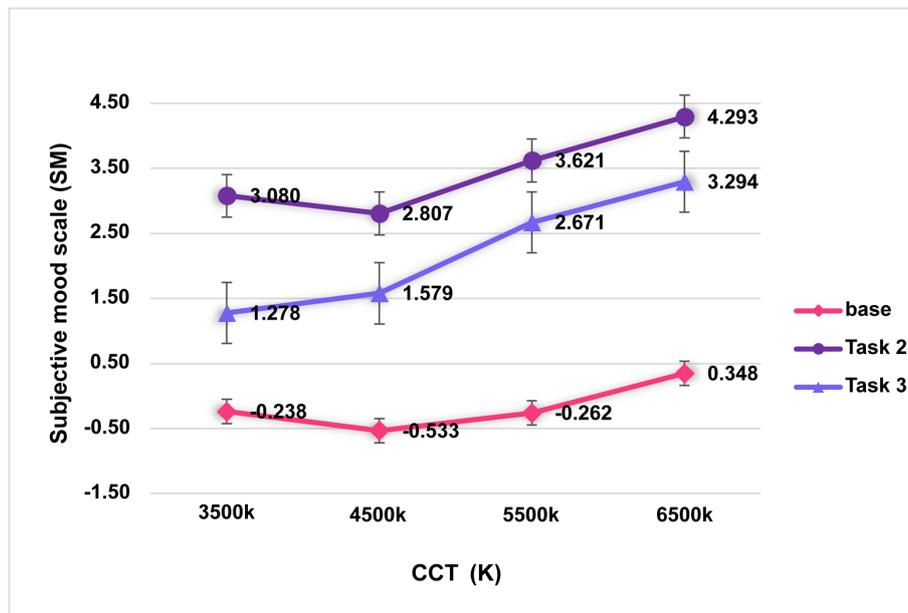

**Fig. 7**. Line chart of mean *SM* at various CCT and task level.

## 5.2 The effect of "CCT x Task" on driver's alterness

According to the EEG findings, the main effect of CCT on driver's alterness was statistically significant but the interaction effect of "CCT x Task" was not. The participants' alertness level significantly increased at medium CCT (4500K) compared to low CCT (3500K), but decreased at 5500K and 6500K (see Fig. 8). In contrast to the findings of Mills et al. (2007) and Viola et al. (2008), this finding indicated that an individual's level of alertness does not rise as the CCT does. We tentatively interpreted the above-mentioned results could be attributed to the following two points: (1) The composition of short-wave blue light in the spectral distribution of middle CCT (4500K) and high CCT (5500K, 6500K) light source is similar, so these three CCT levels have relatively similar stimulation intensity on drivers' ipRGCs, resulting in no significant difference in drivers' alertness level; (2) according to the analysis of the results of the driver's visual comfort and fatigue under various CCTs, it could be known that driving under high CCT for a long time may lead to the deepening of participants' fatigue, resulting in decreases in the level of alterness of participants.

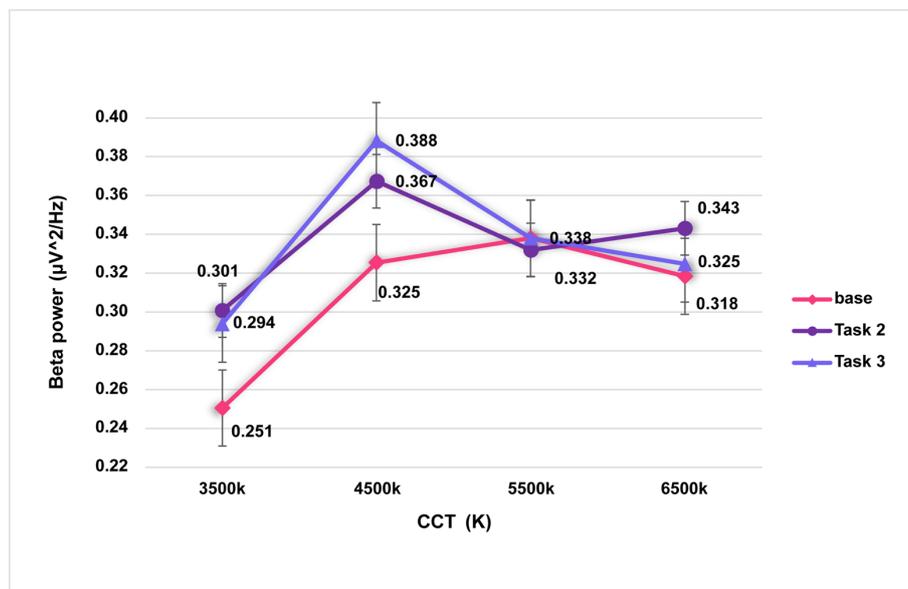

**Fig. 8**. Beta power for alterness at different CCT and task levels.

The interaction of "CCT x Task" had a significant impact on participants' subjective alertness. The results of the simple effect of CCT showed that only in task 1 (base), participants believed that their alertness level was significantly increased in the environment of high CCT (5500K, 6500K) compared with low CCT (3500K). This suggested that high CCT may have an immediate and short-term intervention effect on individual subjective alertness. As the CCT level increased, the subjective alertness level of the subjects also showed a trend of rising first and then decreasing, and the subjects believed that the subjective alertness level was the highest at the middle CCT (4500K) (see Fig. 9). This may be related to the physiological, psychological characteristics and preferences of the subjects who are more inclined to drive in a middle CCT light environment with relatively moderate subjective perception of luminance and CCT level. In addition, the aformentionded phenomenon also implied that appropriately raising the CCT level of the nighttime road lighting environment can aid drivers in maintaining a higher level of subjective alertness, whereas an excessively high CCT will cause the driver to experience visual and psychological discomfort (as evidenced by the findings of visual fatigue and subjective lighting evaluation), on the contrary, it will result in a decline in the driver's subjective alterness and attenuate their attention level, which is in line with the research findings of (Young et al., 2015).

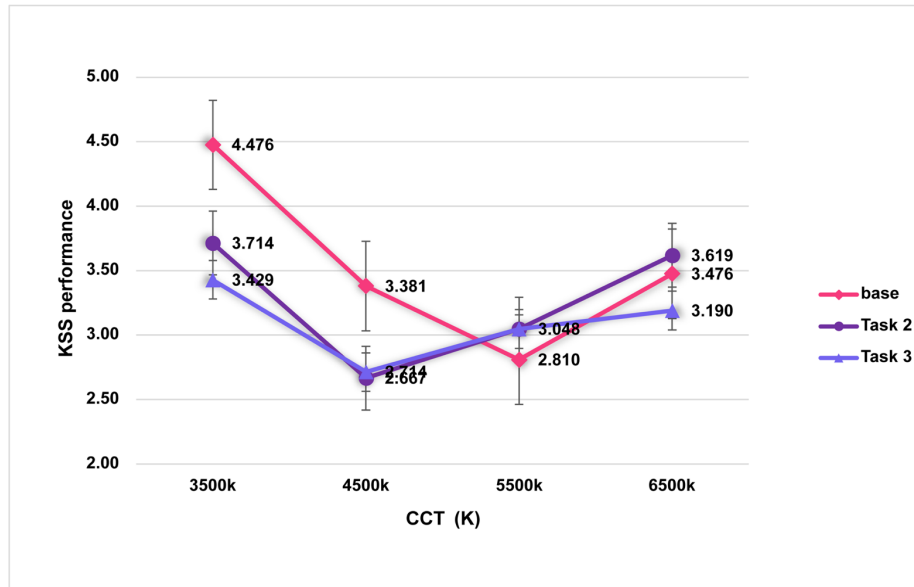

**Fig. 9.** KSS performance for alterness at different CCT and task levels.

## 5.3 The effect of "CCT x Task" on driver's fatigue

According to the analysis results, there was no statistically significant interaction of "CCT x Task" and no main effect of CCT (and Task) on driver's fatigue ($\theta/\beta$), subjective visual fatigue and CFF value. This may be attributed to the short task duration setting (8 min in the current study) and the task load did not reach the average fatigue critical point of the subjects. It was found that with the increase of CCT, the value of $\theta/\beta$ in each task showed a trend of first decreasing and then increasing and then decreasing (see Fig. 10), and the value of $\theta/\beta$ was the lowest at the medium CCT (4500K). This indicated that the medium CCT environment may be the most beneficial to drivers to maintain a better mental state, while the high CCT (5500K) may aggravate the mental fatigue of the subjects. Comparing the values of $\theta/\beta$ at the two high CCTs (5500K vs. 6500K), it was found that the degree of mental fatigue of participants seemed to be lessened at 6500K compared with 5500K. The possible explanation may be that with the increase of CCT, the level of alertness ($\beta$) of participants also increased (6500K > 5500K), but the degree of visual fatigue also increased (6500K > 5500K), too. In the present study, the ratio of theta to beta ($\theta/\beta$) was used to represent the dynamic balance between individual's states of sleepiness and alertness. Therefore, it may lead to the above-mentioned fluctuation characteristics in the degree of mental fatigue of paticipants.

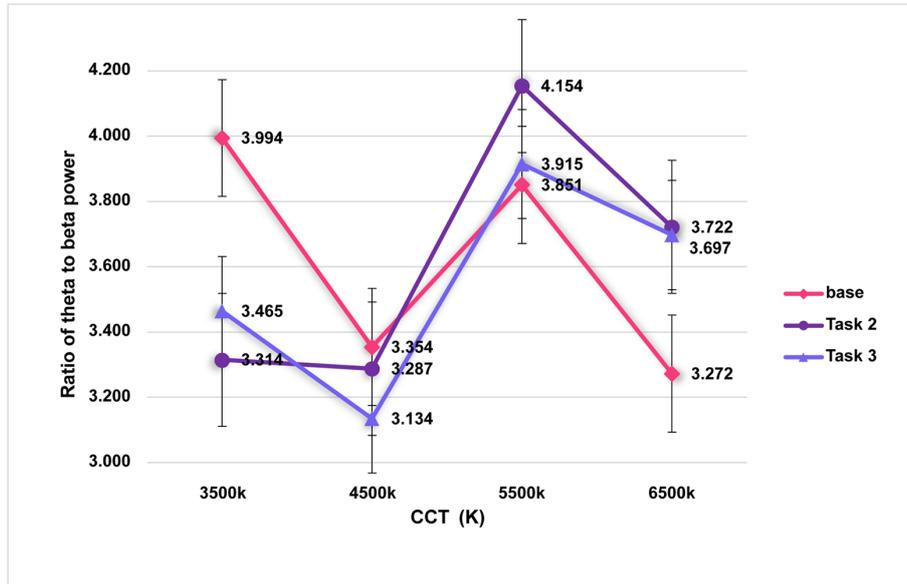

**Fig. 10**. Ratio of theta to beta power for driver's fatigue at different CCT and task level.

As shown in Fig. 11, with the increase of CCT, the subjective visual fatigue score of each task stage generally shows an upward trend, which is quite different from the relevant research conclusions of indoor office lighting. This suggests that individuals' visual adaptation habits and natural biorhythms need to be taken into special consideration when designing nighttime road lighting schemes. Therefore, although increasing the CCT level of road lighting at night can improve the individual's alertness to a certain extent, providing an environment with too high CCT may instead aggravate their subjective visual fatigue. Hence, road lighting designers should make a balance and choose a moderate solution. As shown in Fig. 12, with the increase of CCT, the CFF value of each task showed a trend of first increasing and then decreasing, and the CFF value was the highest at the medium CCT (4500K). This indicated that the medium CCT environment may be the most beneficial for drivers to maintain a better mental state, while the high CCT (about 5500K) may aggravate the mental fatigue of the subjects, which further verified the reliability of the EEG evaluation results.

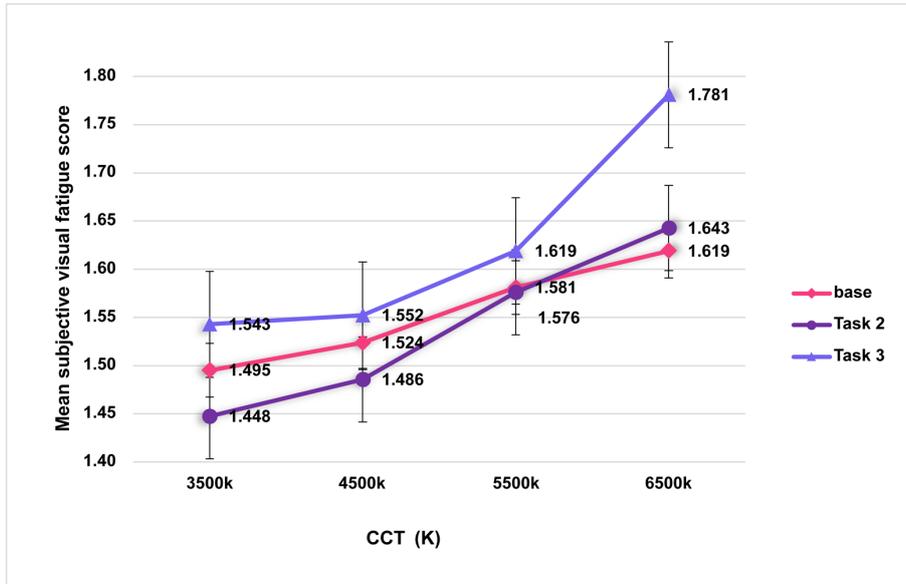

**Fig. 11**. Mean subjective visual fatigue score at different CCT and task levels.

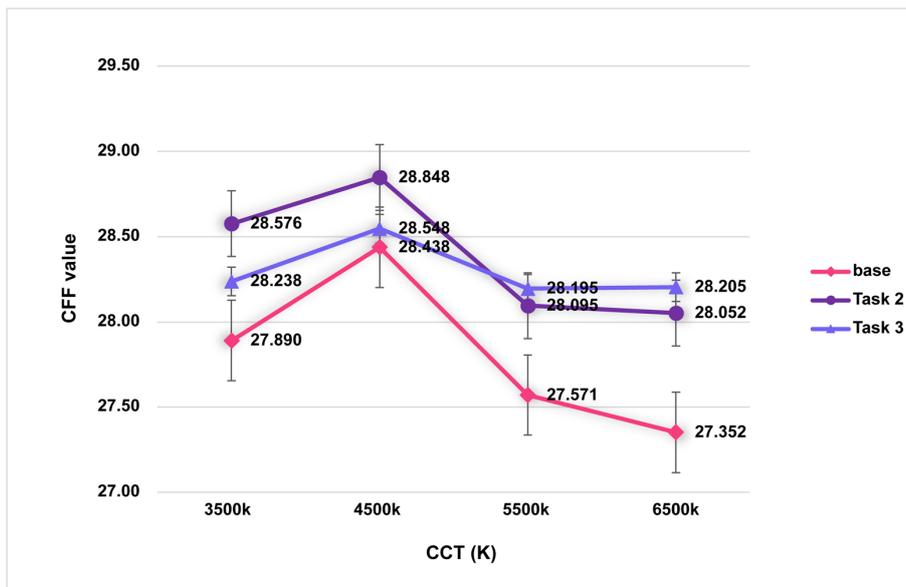

**Fig. 13**. CFF score at different CCT and task levels.

### 5.4 The effect of CCT on driver's reaction time (RT)

The results of RT demonstrated that CCT has a significant impact on the RT of participants (see Fig. 14). Specifically, the RT of the driver decreased overall as the CCT of the external light source increased, which is consistent with the findings of previous relevant studies (Beaven & Ekström, 2013; Chellappa et al., 2011; Okamoto & Nakagawa, 2015). This also implied that there is a relatively high risk associated with low CCT level (3500K). Increasing the CCT in the right way would enhance the driver's responsiveness and safety while driving.

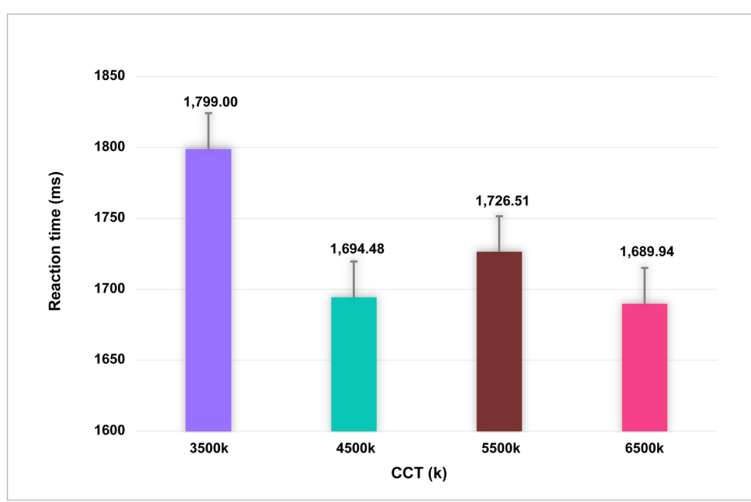

**Fig. 14**. Bar chart of mean reaction time at different CCT levels.

**5.5 Discussion of experimental results of subjective lighting evaluation**

The results of subjective lighting evaluation score demonstrated that participants' subjective comfort and subjective sense of security could be significantly impacted by different CCTs. The participants thought that the high CCTs (5500K, 6500K) were perceived as more glaring than the low CCT (3500K) and middle CCT (4500K) (see Fig. 15 (a)), and the cold white light color caused a lack of psychological security (see Fig. 15 (b)), which was consistent with the findings of Yu and Akita (2019). In terms of subjective color discrimination ability (see Fig. 15 (c)), subjective visual clarity (see Fig. 15 (d)) and subjective satisfaction (see Fig. 15 (e)), there is no discernible difference between individual's subjective feelings under different CCT levels. Additionally, a higher CCT made people feel brighter (see Fig. 15 (f)), which is consistent with common sense. Similar to the findings of Huang et al. (2015), participants' self-evaluation results for visual clarity and color discrimination ability at the medium color temperature (4500K) were superior to those at the other three color temperatures (3500K, 5500K and 6500K). Moreover, the participants's subjective satisfaction with the higher color temperature (6500K) lighting scheme for nighttime driving was the lowest, confirming once again that it was inappropriate to use a high CCT level at night.

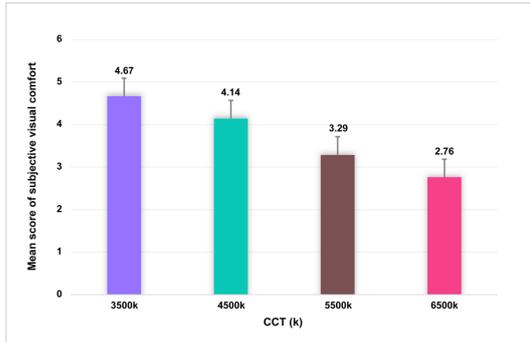

(a) Bar chart of mean subjective visual comfort scores at four CCT levels.

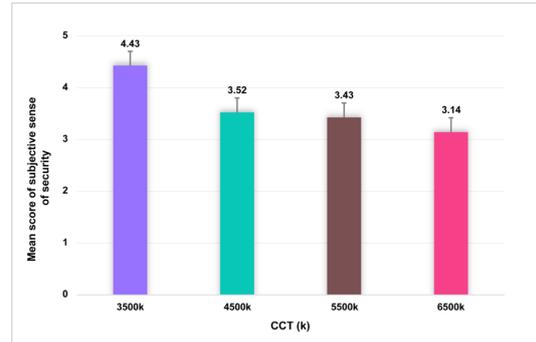

(b) Bar chart of mean scores of subjective feeling of security at four CCT levels.

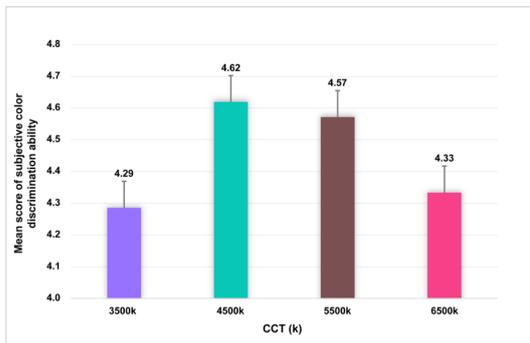

(c) Bar chart of mean score of subjective color discrimination ability at four CCT levels.

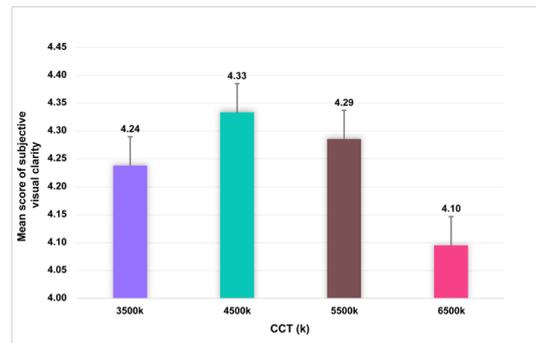

(d) Bar chart of mean score of subjective visual clarity at four CCT levels.

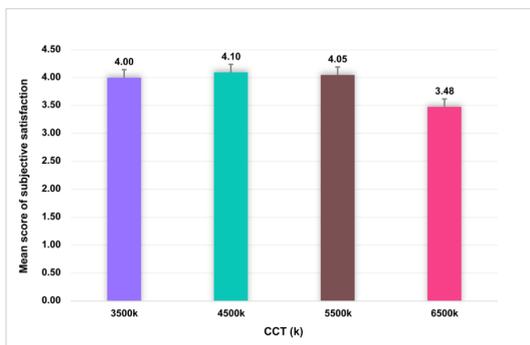

(e) Bar chart of mean score of subjective satisfaction at four CCT levels.

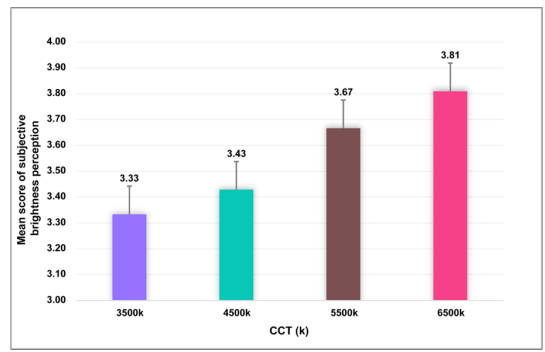

(f) Bar chart of mean score of subjective brightness perception at four CCT levels.

**Fig. 15**. Experimental results of subjective lighting evaluation scores. (a) depicts the mean scores of subjective visual comfort at four CCT levels; (b) depicts the mean scores of subjective sense of security at four CCT levels; (c) depicts the mean scores of subjective color discrimination ability at four CCT levels; (d) depicts the mean scores of subjective visual clarity at four CCT levels; (e) depicts the mean scores of subjective satisfaction at four CCT levels; (f) depicts the mean scores of subjective brightness perception at four CCT levels (3500K, 4500K, 5500K and 6500K).

# 6. Limitation

The current study aimed to investigate the influence of different CCT levels on non-visual indicators (e.g., mood, alertness, fatigue, and reaction time) of individuals performing different driving activities by using the triangulation method of "EEG evaluation + subjective evaluation + behavioral evaluation".

Although different behavioural, cognitive responses and electrophysiological indicators, as measured by EEG, subejctive evaluation and behavioural responses, were observed and discussed, future research will consider the following.

First, in the experimental design, only the CCT was included, while the illuminance that exists in real situations and has an interrelationship with CCT was ignored. Thus, in the future, it is necessary to study the regulating effect of illuminance factor on drivers' non-visual effects, and further carry out experimental research on the non-visual effects of the interaction of "CCT * Illuminance" on drivers. Second, the experiment was conducted in a simulated laboratory setting and the results can not be directly generalized to practical application scenarios, and if possible, real nighttime road test will be carried out in the future. Third, the time of driving task was set 8 min in the present study. The time limit is too short, and the non-visual effect may not be sufficiently induced. Thus, we can extend the time of three types of tasks. Fourth, in the actual nighttime driving light environment, in addition to road lighting, there is also interference from other additional light sources such as headlights, in-car center-control display light source, and so on. Therefore, in the follow-up research, on the one hand, the technical scheme of indoor simulation platform construction can be further improved to simulate and restore the actual nighttime urban road lighting environment to the greatest extent, thereby saving research costs; On the other hand, we can try to develop and build an outdoor closed road lighting control system to meet the needs of real road testing, verify the research results in the laboratory, and improve the reliability and validity of the research results. Finally, this study selected three typical driving tasks to explore the non-visual effects of different color temperature and illumination conditions on drivers (non-professional drivers) during a fixed time period at night (19:00-20:30), but the tasks involved in the actual driving situation are more complex. Furthermore, non-visual effects of light influencing factors also include light time (e.g., season, day, week, month and year), light duration, light spatial distribution, and biological rhythm system; additionally, non-visual effects of individual influencing factors also include occupation, gender, age, and so on. Therefore, this research is only a preliminary investigation into the non-visual effects of road lighting, and there are still rich research contents that can be further enhanced and further expanded in the future.

## 7. Conclusion

In conclusion, according to the above analysis, the significant interaction of "CCT x Task" is mainly reflected in the aspect of individual alertness and reaction time. Individual's subjective emotional experience, subjective visual comfort and subjective feelings of security are more sensitive to the changes in the CCT of external road lighting environment at night, whereas their mental fatigue and visual fatigue are less affected by the CCT of external light environment. The results of the subjective and objective comprehensive evaluation analysis demonstrated that the EEG evaluation indexes selected fro this study could accurately reflect the characteristics of various non-visual indicators (mood, alertness, fatigue and reaction time) under different experimental levels. The results of various non-visual indicators at the CCT level also suggested that:

(1) In the spectrum of light source with low CCT (3500K), the energy of long-wave red light is higher, which is more conducive to the adaptation of individual dark vision in the night environment and conforms to the natural visual physiological habits of human body. Furthermore, the warm color temperature atmosphere gives people a sense of warmth and security, so the subjective visual comfort evaluation of drivers under this CCT is the best, and the mean score of subjective satisfaction of drivers is relatively high. Low CCT environments are also more likely to make drivers fall into the "comfort zone", particularly when performing monotonous driving tasks, which can be more likely to cause fatigue and sleepiness, as well as prolong the driver's reaction time in an extent and increase the potential driving risk.

(2) Middle CCT (4500K) appears to be the most beneficial for drivers to maintain an ideal physical and mental state during nighttime driving, as manifested by: good mood experience (mainly fluctuated in the range of neutral and positive mood); it is conducive to the driver maintaining a relatively stable and high level of alertness, and can respond quickly to external stimuli; the degree of mental fatigue and visual fatigue is relatively low.

(3) Although the reaction time of the participants was faster in the high CCT environment (> 5500K), the environment was more likely to aggravate the mental fatigue of the individuals during the long driving process, induce the negative emotions of the individuals, cause the eye discomfort, and cause the drivers lack of psychological security. Participants were also dissatisfied with the lighting scheme's excessively high CCT (6500K). Therefore, it is tentatively concluded that the middle CCT (around 4500K) is more suitable or appropriate for the nighttime road lighting scheme, and the light with cold white temperature (> 5500K) should not be provided in excess. The aforementioned experimental findings can also serve as a foundation for adjusting and defining variables and parameter ranges for the subsequent simulation experiment of non-visual effect on drivers caused by the interaction of CCT, illuminance and driving task.

## Data Availability Statement

The data used to support the findings of this study are available from the corresponding author upon request.

# Reference

Almeida, P. S., Bender, V. C., Braga, H. A., Dalla Costa, M. A., Marchesan, T. B., & Alonso, J. M. (2014). Static and dynamic photoelectrothermal modeling of LED lamps including low-frequency current ripple effects. IEEE Transactions on Power Electronics, 30(7), 3841–3851. https://doi.org/10.1109/TPEL.2014.2340352

Bailes, H. J., & Lucas, R. J. (2013). Human melanopsin forms a pigment maximally sensitive to blue light (λ max≈ 479 nm) supporting activation of Gq/11 and Gi/o signalling cascades. Proceedings of the Royal Society B: Biological Sciences, 280(1759), 20122987. https://doi.org/10.1098/rspb.2012.2987

Baumann, M., & Krems, J. F. (2007). Situation awareness and driving: A cognitive model. Modelling Driver Behaviour in Automotive Environments: Critical Issues in Driver Interactions with Intelligent Transport Systems, 253–265.

Beaven, C. M., & Ekström, J. (2013). A comparison of blue light and caffeine effects on cognitive function and alertness in humans. PloS One, 8(10), e76707. https://doi.org/10.1371/journal.pone.0076707

Bellia, L., Bisegna, F., & Spada, G. (2011). Lighting in indoor environments: Visual and non-visual effects of light sources with different spectral power distributions. Building and Environment, 46(10), 1984–1992. https://doi.org/10.1016/j.buildenv.2011.04.007

Berka, C., Levendowski, D. J., Lumicao, M. N., Yau, A., Davis, G., Zivkovic, V. T., Olmstead, R. E., Tremoulet, P. D., & Craven, P. L. (2007). EEG correlates of task engagement and mental workload in vigilance, learning, and memory tasks. Aviation, Space, and Environmental Medicine, 78(5), B231–B244.

Bose, R., Wang, H., Dragomir, A., Thakor, N. V., Bezerianos, A., & Li, J. (2019). Regression-based continuous driving fatigue estimation: Toward practical implementation. IEEE Transactions on Cognitive and Developmental Systems, 12(2), 323–331. https://doi.org/10.1109/TCDS.2019.2929858

Brainard, G. C., Hanifin, J. P., Greeson, J. M., Byrne, B., Glickman, G., Gerner, E., & Rollag, M. D. (2001). Action spectrum for melatonin regulation in humans: Evidence for a novel circadian photoreceptor. Journal of Neuroscience, 21(16), 6405–6412. https://doi.org/10.1523/JNEUROSCI.21-16-06405.2001

Brown, L., Grundlehner, B., & Penders, J. (2011). Towards wireless emotional valence detection from EEG. 2011 Annual International Conference of the IEEE Engineering in Medicine and Biology Society, 2188–2191. https://doi.org/10.1109/IEMBS.2011.6090412

Bullough, J. D., & Rea, M. S. (2004). Visual performance under mesopic conditions: Consequences for roadway lighting. Transportation Research Record, 1862(1), 89–94. https://doi.org/10.3141/1862-11

Cajochen, C., Munch, M., Kobialka, S., Krauchi, K., Steiner, R., Oelhafen, P., Orgul, S., & Wirz-Justice, A. (2005). High sensitivity of human melatonin, alertness, thermoregulation, and heart rate to short wavelength light. The Journal of Clinical Endocrinology & Metabolism, 90(3), 1311–1316. https://doi.org/10.1210/jc.2004-0957

Cajochen, C., Zeitzer, J. M., Czeisler, C. A., & Dijk, D.-J. (2000). Dose-response relationship for light intensity and ocular and electroencephalographic correlates of human alertness. Behavioural Brain Research, 115(1), 75–83. https://doi.org/10.1016/S0166-4328(00)00236-9

Chellappa, S. L., Steiner, R., Blattner, P., Oelhafen, P., Götz, T., & Cajochen, C. (2011). Non-visual effects of light on melatonin, alertness and cognitive performance: Can blue-enriched light keep us alert? PloS One, 6(1), e16429. https://doi.org/10.1371/journal.pone.0016429

Cunningham, M. L., & Regan, M. A. (2016). The impact of emotion, life stress and mental health issues on driving performance and safety. Road & Transport Research: A Journal of Australian and New Zealand Research and Practice, 25(3), 40–50.

Davidovic, M., Djokic, L., Cabarkapa, A., Djuretic, A., Skerovic, V., & Kostic, M. (2019). Drivers'

preference for the color of LED street lighting. IEEE Access : Practical Innovations, Open Solutions, 7, 72850–72861.

Delorme, A., & Makeig, S. (2004). EEGLAB: An open source toolbox for analysis of single-trial EEG dynamics including independent component analysis. Journal of Neuroscience Methods, 134(1), 9–21. https://doi.org/10.1016/j.jneumeth.2003.10.009

Domenichini, L., La Torre, F., Vangi, D., Virga, A., & Branzi, V. (2017). Influence of the lighting system on the driver's behavior in road tunnels: A driving simulator study. Journal of Transportation Safety & Security, 9(2), 216–238. https://doi.org/10.1080/19439962.2016.1173155

Dong, L., Qin, L., Xu, W., & Zhang, L. (2017). The impact of LED correlated color temperature on visual performance under mesopic conditions. IEEE Photonics Journal, 9(6), 1–16.

Dong, L., Shang, X., Zhao, Y., Qin, L., & Xu, W. (2018). The impact of LED light color on the dark adaptation of human vision in tunnel entrances. IEEE Photonics Journal, 10(5), 1–11.

Dong, L., Zhao, E., Chen, Y., Qin, G., & Xu, W. (2020). Impact of LED color temperatures on perception luminance in the interior zone of a tunnel considering fog transmittance. Advances in Civil Engineering, 2020, 1–13.

Elvik, R. (1995). Meta-analysis of evaluations of public lighting as accident countermeasure. Transportation Research Record, 1485(1), 12–24.

Englezou, M., & Michael, A. (2022). Evaluation of visual and non-visual effects of daylighting in healthcare patient rooms using climate-based daylight metrics and melanopic metrics. E3S Web of Conferences, 362, 01003. https://doi.org/10.1051/e3sconf/202236201003

Faul, F., Erdfelder, E., Lang, A.-G., & Buchner, A. (2007). G* Power 3: A flexible statistical power analysis program for the social, behavioral, and biomedical sciences. Behavior Research Methods, 39(2), 175–191. https://doi.org/10.3758/BF03193146

Figueiro, M. G., Sahin, L., Wood, B., & Plitnick, B. (2016). Light at night and measures of alertness and performance: Implications for shift workers. Biological Research for Nursing, 18(1), 90–100. https://doi.org/10.1177/1099800415572873

Flynn, J. E., Spencer, T. J., Martyniuk, O., & Hendrick, C. (1973). Interim study of procedures for investigating the effect of light on impression and behavior. Journal of the Illuminating Engineering Society, 3(1), 87–94. https://doi.org/10.1080/00994480.1973.10732231

Fotios, S., & Cheal, C. (2007). Lighting for subsidiary streets: Investigation of lamps of different SPD. Part 2—Brightness. Lighting Research & Technology, 39(3), 233–249.

Hathaway, W. E. (1993). Non-visual effects of classroom lighting on children. Education Canada, 33(4), 34–40.

He, S., Liang, B., Tähkämö, L., Maksimainen, M., & Halonen, L. (2020). The influences of tunnel lighting environment on drivers' peripheral visual performance during transient adaptation. Displays, 64, 101964. https://doi.org/10.1016/j.displa.2020.101964

He, S., Tähkämö, L., Maksimainen, M., Liang, B., Pan, G., & Halonen, L. (2017). Effects of transient adaptation on drivers' visual performance in road tunnel lighting. Tunnelling and Underground Space Technology, 70, 42–54. https://doi.org/10.1016/j.tust.2017.07.008

Herrmann, W., Fichte, K., Freund, G., & others. (1979). Reflections on the topics: EEG frequency bands and regulation of vigilance. Pharmacopsychiatry, 12(02), 237–245. https://doi.org/10.1055/s-0028-1094615

Hoseinabadi, S., Porabdeyan, S., Zare, M., Amiri, S., Ghasemi, M., & Mansori, A. (2015). Does traffic stress affect distance estimation and recognition accuracy in urban bus drivers? Archives of Environmental & Occupational Health, 70(4), 214–217. https://doi.org/10.1080/19338244.2013.859121

Huang, R.-H., Lee, L., Chiu, Y.-A., & Sun, Y. (2015). Effects of correlated color temperature on focused and sustained attention under white LED desk lighting. Color Research & Application, 40(3), 281–286. https://doi.org/10.1002/col.21885

Iskra-Golec, I., Wazna, A., & Smith, L. (2012). Effects of blue-enriched light on the daily course of mood, sleepiness and light perception: A field experiment. Lighting Research & Technology, 44(4),

506–513. https://doi.org/10.1177/1477153512447528

Jackson, A. F., & Bolger, D. J. (2014). The neurophysiological bases of EEG and EEG measurement: A review for the rest of us. Psychophysiology, 51(11), 1061–1071. https://doi.org/10.1111/psyp.12283

Jap, B. T., Lal, S., Fischer, P., & Bekiaris, E. (2009). Using EEG spectral components to assess algorithms for detecting fatigue. Expert Systems with Applications, 36(2), 2352–2359. https://doi.org/10.1016/j.eswa.2007.12.043

Kang, C., Li, W., Wang, Z., Li, S., Huang, Z., & Wu, K. (2021). Influence of LED color temperature on visual performance during the process of dark adaptation in tunnels. Journal of South China University of Technology, 49(4), 117–123.

KHADEMAGHA, P., DIEPENS, J. F., ARIES, M. B., ROSEMANN, A. L., & VAN LOENEN, E. J. (2015). Effect of different design parameters on the visual and non-visual assessment criteria in office spaces. Proceedings of International Conference CISBAT 2015 Future Buildings and Districts Sustainability from Nano to Urban Scale, Article CONF.

Khushaba, R. N., Kodagoda, S., Lal, S., & Dissanayake, G. (2010). Driver drowsiness classification using fuzzy wavelet-packet-based feature-extraction algorithm. IEEE Transactions on Biomedical Engineering, 58(1), 121–131. https://doi.org/10.1109/TBME.2010.2077291

Krems, J. F., & Baumann, M. R. (2009). Driving and situation awareness: A cognitive model of memory-update processes. Human Centered Design: First International Conference, HCD 2009, Held as Part of HCI International 2009, San Diego, CA, USA, July 19-24, 2009 Proceedings 1, 986–994. https://doi.org/10.1007/978-3-642-02806-9_113

Kuze, J., & Ukai, K. (2008). Subjective evaluation of visual fatigue caused by motion images. Displays, 29(2), 159–166. https://doi.org/10.1016/j.displa.2007.09.007

Lal, S. K., & Craig, A. (2001). A critical review of the psychophysiology of driver fatigue. Biological Psychology, 55(3), 173–194. https://doi.org/10.1016/S0301-0511(00)00085-5

Landis, C. (1954). Determinants of the critical flicker-fusion threshold. Physiological Reviews, 34(2), 259–286. https://doi.org/10.1152/physrev.1954.34.2.259

Lang, P. J. (2005). International affective picture system (IAPS): Affective ratings of pictures and instruction manual. Technical Report.

Lee, B.-G., Lee, B.-L., & Chung, W.-Y. (2014). Mobile healthcare for automatic driving sleep-onset detection using wavelet-based EEG and respiration signals. Sensors, 14(10), 17915–17936. https://doi.org/10.3390/s141017915

Lewis, A. L. (1999). Visual performance as a function of spectral power distribution of light sources at luminances used for general outdoor lighting. JOURNAL of the Illuminating Engineering Society, 28(1), 37–42. https://doi.org/10.1080/00994480.1999.10748250

Li, H., Wang, H., Shen, J., Sun, P., Xie, T., Zhang, S., & Zheng, Z. (2017). Non-visual biological effects of light on human cognition, alertness, and mood. Light in Nature VI, 10367, 56–61. https://doi.org/10.1117/12.2272555

Li, W., He, Q., Fan, X., & Fei, Z. (2012). Evaluation of driver fatigue on two channels of EEG data. Neuroscience Letters, 506(2), 235–239. https://doi.org/10.1016/j.neulet.2011.11.014

Li, X., Ling, J., Shen, Y., Lu, T., Feng, S., & Zhu, H. (2021). The impact of CCT on driving safety in the normal and accident situation: A VR-based experimental study. Advanced Engineering Informatics, 50, 101379. https://doi.org/10.1016/j.aei.2021.101379

Liang, B., He, S., Tähkämö, L., Tetri, E., Cui, L., Dangol, R., & Halonen, L. (2020). Lighting for road tunnels: The influence of CCT of light sources on reaction time. Displays, 61, 101931. https://doi.org/10.1016/j.displa.2019.101931

Liang, S.-F., Lin, C., Wu, R., Chen, Y., Huang, T., & Jung, T. (2006). Monitoring driver's alertness based on the driving performance estimation and the EEG power spectrum analysis. 2005 IEEE Engineering in Medicine and Biology 27th Annual Conference, 5738–5741.

Lin, C.-T., Wu, R.-C., Jung, T.-P., Liang, S.-F., & Huang, T.-Y. (2005). Estimating driving performance based on EEG spectrum analysis. EURASIP Journal on Advances in Signal Processing, 2005, 1–10. https://doi.org/10.1155/ASP.2005.3165

Liu, T. (2020). Lighting evaluation and design for the stockholm metro system based on current models for non-visual responses.

Liu, Y., Peng, L., Lin, L., Chen, Z., Weng, J., & Zhang, Q. (2021). The impact of LED spectrum and correlated color temperature on driving safety in long tunnel lighting. Tunnelling and Underground Space Technology, 112, 103867. https://doi.org/10.1016/j.tust.2021.103867

Lockley, S. W., Evans, E. E., Scheer, F. A., Brainard, G. C., Czeisler, C. A., & Aeschbach, D. (2006). Short-wavelength sensitivity for the direct effects of light on alertness, vigilance, and the waking electroencephalogram in humans. Sleep, 29(2), 161–168.

Makeig, S., & Inlow, M. (1993). Lapse in alertness: Coherence of fluctuations in performance and EEG spectrum. Electroencephalography and Clinical Neurophysiology, 86(1), 23–35. https://doi.org/10.1016/0013-4694(93)90064-3

Mayeur, A., Bremond, R., & Bastien, J. C. (2010). The effect of the driving activity on target detection as a function of the visibility level: Implications for road lighting. Transportation Research Part F: Traffic Psychology and Behaviour, 13(2), 115–128. https://doi.org/10.1016/j.trf.2009.12.004

Meng, X., Xiao-dong, P., Feng, C., & Xiao-xiang, M. (2020). Experimental study on the effect of color light vision regulation on hypnotic relief of long tunnel driving. China Journal of Highway and Transport, 33(11), 235.

Mills, P. R., Tomkins, S. C., & Schlangen, L. J. (2007). The effect of high correlated colour temperature office lighting on employee wellbeing and work performance. Journal of Circadian Rhythms, 5(1), 1–9.

Morales, J. M., Díaz-Piedra, C., Rieiro, H., Roca-González, J., Romero, S., Catena, A., Fuentes, L. J., & Di Stasi, L. L. (2017). Monitoring driver fatigue using a single-channel electroencephalographic device: A validation study by gaze-based, driving performance, and subjective data. Accident Analysis & Prevention, 109, 62–69. https://doi.org/10.1016/j.aap.2017.09.025

Nemeth, C. J. (1995). Dissent as driving cognition, attitudes, and judgments. Social Cognition, 13(3), 273–291. https://doi.org/10.1521/soco.1995.13.3.273

Okamoto, Y., & Nakagawa, S. (2015). Effects of daytime light exposure on cognitive brain activity as measured by the ERP P300. Physiology & Behavior, 138, 313–318. https://doi.org/10.1016/j.physbeh.2014.10.013

Organization, W. H. (2018). Global status report on road safety 2018. World Health Organization.

Pachito, D. V., Eckeli, A. L., Desouky, A. S., Corbett, M. A., Partonen, T., Rajaratnam, S. M., & Riera, R. (2018). Workplace lighting for improving alertness and mood in daytime workers. Cochrane Database of Systematic Reviews, 3. https://doi.org/10.1002/14651858.CD012243.pub2

Palmiero, M., & Piccardi, L. (2017). Frontal EEG asymmetry of mood: A mini-review. Frontiers in Behavioral Neuroscience, 11, 224. https://doi.org/10.3389/fnbeh.2017.00224

Pan, Z., Wang, H., Wu, J., & Chen, Q. (2023). Non-visual effects of CCT on drivers, evidence from EEG. Intelligent Human Systems Integration 2023, 69, 1.

Peng, L., Weng, J., Yang, Y., & Wen, H. (2022). Impact of light environment on driver's physiology and psychology in interior zone of long tunnel. Frontiers in Public Health, 10. https://doi.org/10.3389/fpubh.2022.842750

Phipps-Nelson, J., Redman, J. R., Schlangen, L. J., & Rajaratnam, S. M. (2009). Blue light exposure reduces objective measures of sleepiness during prolonged nighttime performance testing. Chronobiology International, 26(5), 891–912. https://doi.org/10.1080/07420520903044364

Plainis, S., Murray, I., & Pallikaris, I. (2006). Road traffic casualties: Understanding the night-time death toll. Injury Prevention, 12(2), 125–138. https://doi.org/10.1136/ip.2005.011056

Pulat, B. M. (1997). Fundamentals of industrial ergonomics. Waveland PressInc.

Rahm, J., & Johansson, M. (2018). Assessing the pedestrian response to urban outdoor lighting: A full-scale laboratory study. PLoS One, 13(10), e0204638. https://doi.org/10.1371/journal.pone.0204638

Razak, S. F. A., Yogarayan, S., Aziz, A. A., Abdullah, M. F. A., & Kamis, N. H. (2022). Physiological-based driver monitoring systems: A scoping review. Civil Engineering Journal, 8(12), 3952–3967. https://doi.org/10.28991/CEJ-2022-08-12-020

Rosenthal, N. E., & Wehr, T. A. (1992). Towards understanding the mechanism of action of light in seasonal affective disorder. Pharmacopsychiatry, 25(01), 56–60. https://doi.org/10.1055/s-2007-1014389

Ru, T., de Kort, Y. A., Smolders, K. C., Chen, Q., & Zhou, G. (2019). Non-image forming effects of illuminance and correlated color temperature of office light on alertness, mood, and performance across cognitive domains. Building and Environment, 149, 253–263. https://doi.org/10.1016/j.buildenv.2018.12.002

Saraiji, R. (2009). A methodology for locating the maximum vertical illuminance in street lighting. Leukos, 6(2), 139–152. https://doi.org/10.1582/LEUKOS.2009.06.02004

Schreuder, D. A. (1998). Road lighting for safety. Thomas Telford.

Scott-Parker, B. (2017). Emotions, behaviour, and the adolescent driver: A literature review. Transportation Research Part F: Traffic Psychology and Behaviour, 50, 1–37.

Shahid, A., Wilkinson, K., Marcu, S., & Shapiro, C. M. (2012). Karolinska sleepiness scale (KSS). STOP, THAT and One Hundred Other Sleep Scales, 209–210.

Skrandies, W. (1985). Critical flicker fusion and double flash discrimination in different parts of the visual field. International Journal of Neuroscience, 25(3–4), 225–231. https://doi.org/10.3109/00207458508985374

Smolders, K. C., De Kort, Y. A., & Cluitmans, P. (2012). A higher illuminance induces alertness even during office hours: Findings on subjective measures, task performance and heart rate measures. Physiology & Behavior, 107(1), 7–16. https://doi.org/10.1016/j.physbeh.2012.04.028

Steinhauser, K., Leist, F., Maier, K., Michel, V., Pärsch, N., Rigley, P., Wurm, F., & Steinhauser, M. (2018). Effects of emotions on driving behavior. Transportation Research Part F: Traffic Psychology and Behaviour, 59, 150–163. https://doi.org/10.1016/j.trf.2018.08.012

Stockman, A., & Sharpe, L. T. (2006). Into the twilight zone: The complexities of mesopic vision and luminous efficiency. Ophthalmic and Physiological Optics, 26(3), 225–239. https://doi.org/10.1111/j.1475-1313.2006.00325.x

Sun, S., Hu, J., & Wang, R. (2021). Correlation between visibility and traffic safety visual distance in foggy areas during the daytime. Traffic Injury Prevention, 22(7), 514–518. https://doi.org/10.1080/15389588.2021.1916924

Thapan, K., Arendt, J., & Skene, D. J. (2001). An action spectrum for melatonin suppression: Evidence for a novel non-rod, non-cone photoreceptor system in humans. The Journal of Physiology, 535(1), 261–267. https://doi.org/10.1111/j.1469-7793.2001.t01-1-00261.x

Van Bommel, W. J. (2006). Non-visual biological effect of lighting and the practical meaning for lighting for work. Applied Ergonomics, 37(4), 461–466. https://doi.org/10.1016/j.apergo.2006.04.009

Vetter, C., Juda, M., Lang, D., Wojtysiak, A., & Roenneberg, T. (2011). Blue-enriched office light competes with natural light as a zeitgeber. Scandinavian Journal of Work, Environment & Health, 437–445. https://doi.org/10.5271/sjweh.3144

Vicente, E. G., Matesanz, B. M., Rodriguez-Rosa, M., Saez, A. M., Mar, S., & Arranz, I. (2023). Effect of correlated color temperature and S/P-ratio of LED light sources on reaction time in off-axis vision and mesopic lighting levels. Leukos, 19(1), 4–15. https://doi.org/10.1080/15502724.2021.1970580

Viola, A. U., James, L. M., Schlangen, L. J., & Dijk, D.-J. (2008). Blue-enriched white light in the workplace improves self-reported alertness, performance and sleep quality. Scandinavian Journal of Work, Environment & Health, 297–306. https://doi.org/10.5271/sjweh.1268

Wang, Q., Xu, H., Gong, R., & Cai, J. (2015). Investigation of visual fatigue under LED lighting based on reading task. Optik, 126(15–16), 1433–1438. https://doi.org/10.1016/j.ijleo.2015.04.033

Webb, A. R. (2006). Considerations for lighting in the built environment: Non-visual effects of light. Energy and Buildings, 38(7), 721–727. https://doi.org/10.1016/j.enbuild.2006.03.004

Whillans, M., & Allen, M. J. (1992). Color defective drivers and safety. Optometry and Vision Science, 69(6), 463–466. https://doi.org/10.1097/00006324-199206000-00009

Wu, J., Du, X., Tong, M., Guo, Q., Shao, J., Chabebe, A., & Xue, C. (2023). Neural mechanisms behind semantic congruity of construction safety signs: An EEG investigation on construction workers.

Human Factors and Ergonomics in Manufacturing & Service Industries, 33(3), 229–245. https://doi.org/10.1002/hfm.20979

Xiao, H., Cai, H., & Li, X. (2021). Non-visual effects of indoor light environment on humans: A review☆. Physiology & Behavior, 228, 113195. https://doi.org/10.1016/j.physbeh.2020.113195

Yang, L., Ma, R., Zhang, H. M., Guan, W., & Jiang, S. (2018). Driving behavior recognition using EEG data from a simulated car-following experiment. Accident Analysis & Prevention, 116, 30–40. https://doi.org/10.1016/j.aap.2017.11.010

Yang, S., Liu, K., & Zhang, H. (2023). Effects of color temperature and lighting mode on visual and psychological perception of pedestrians in subway station. Second International Conference on Electronic Information Engineering and Computer Communication (EIECC 2022), 12594, 8–15. https://doi.org/10.1117/12.2671360

Yi, H., Changbin, L., Aiguo, W., & Shouzhong, F. (2012). LED lighting control system in tunnel based on intelligent illumination curve. 2012 Fifth International Conference on Intelligent Computation Technology and Automation, 698–701. https://doi.org/10.1109/ICICTA.2012.185

Yong, Y., Zuojun, B., Chuanzheng, Z., & Lei, W. (2011). Study on the mesopic vision theory used in road tunnel lighting measurement. 2011 Third International Conference on Measuring Technology and Mechatronics Automation, 3, 565–567. https://doi.org/10.1109/ICMTMA.2011.711

Yoshiike, T., Honma, M., Yamada, N., Kim, Y., & Kuriyama, K. (2018). Effects of bright light exposure on human fear conditioning, extinction, and associated prefrontal activation. Physiology & Behavior, 194, 268–276. https://doi.org/10.1016/j.physbeh.2018.06.015

Young, C. R., Jones, G. E., Figueiro, M. G., Soutière, S. E., Keller, M. W., Richardson, A. M., Lehmann, B. J., & Rea, M. S. (2015). At-sea trial of 24-h-based submarine watchstanding schedules with high and low correlated color temperature light sources. Journal of Biological Rhythms, 30(2), 144–154. https://doi.org/10.1177/0748730415575432

Yu, H., & Akita, T. (2019). The effect of illuminance and correlated colour temperature on perceived comfort according to reading behaviour in a capsule hotel. Building and Environment, 148, 384–393. https://doi.org/10.1016/j.buildenv.2018.11.027

Zeng, Y., Sun, H., Lin, B., & Zhang, Q. (2021). Non-visual effects of office light environment: Field evaluation, model comparison, and spectral analysis. Building and Environment, 197, 107859. https://doi.org/10.1016/j.buildenv.2021.107859

Zhang, C., & Eskandarian, A. (2020). A survey and tutorial of EEG-based brain monitoring for driver state analysis. ArXiv Preprint ArXiv:2008.11226.

Zhang, Q., Chen, Z., Hu, Y., & Yang, C. (2008). Study on the influence of lighting source color temperature on visual performance in tunnel and road lighting. China Illuminating Engineering Journal, 19(2), 24–29.

Zhang, X., Hu, J., Wang, R., Gao, X., & He, L. (2017). The comprehensive efficiency analysis of tunnel lighting based on visual performance. Advances in Mechanical Engineering, 9(4), 1687814017696449. https://doi.org/10.1177/1687814017696449